# LANet: An Enriched Knowledgebase for Location-aware Activity Recommendation System

A DISSERTATION

submitted towards the fulfillment of the
requirements for the award of the degree of

Master of Technology
*in*
Computer Science and Engineering

By

SAHISNU MAZUMDER

Under the Supervision of
Dr. Dhaval Patel, IIT Roorkee

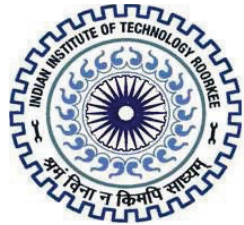

Department of Computer Science and Engineering
INDIAN INSTITUTE OF TECHNOLOGY- ROORKEE
ROORKEE−247667, INDIA.

JUNE 2014

# CANDIDATE'S DECLARATION

I declare that the work presented in this dissertation with title "**LANet: An Enriched Knowledgebase for Location-aware Activity Recommendation System**" towards the fulfilment of the requirement for the award of the degree of **Master of Technology** in **Computer Science & Engineering** submitted in the **Dept. of Computer Science & Engineering**, **Indian Institute of Technology- Roorkee**, India is an authentic record of my own work carried out during the period **from June 2013 to June 2014** under the supervision of **Dr. Dhaval Patel**, Assistant Professor, Dept. of CSE, IIT Roorkee.

The content of this dissertation has not been submitted by me for the award of any other degree of this or any other institute.

DATE : ……………………    SIGNED: ……………………………………
PLACE: ……………………                 (SAHISNU MAZUMDER)

i

# CERTIFICATE

This is to certify that the statement made by the candidate is correct to the best of my knowledge and belief.

DATE : ..........................       SIGNED: ........................................

(DR. DHAVAL PATEL)

ASSISTANT PROFESSOR

DEPT. OF CSE, IIT ROORKEE

*"My big thesis is that although the world looks messy and chaotic, if you translate it into the world of numbers and shapes, patterns emerge and you start to understand why things are the way they are."*

**- Marcus du Sautoy.**

Part of the work presented in this dissertation has been published in the following conference-

Sahisnu Mazumder, Dhaval Patel and Sameep Mehta. **ActMiner: Discovering Location-Specific Activities from Community-Authored Reviews**, In $16^{th}$ International Conference on *Data Warehousing and Knowledge Discovery* (**DaWaK 2014**), (pp. 332-344), Springer International Publishing.

# ABSTRACT

ACCUMULATION of large amount of location-specific reviews on web due to escalating popularity of *Location-based Social Networking* platforms like *Yelp*, *Foursquare*, *Brightkite* etc. in recent years, has created the opportunity to discover location-specific activities and develop myriads of location-aware activity recommendation applications. The performance and popularity of such recommendation applications greatly depend on the *richness* and *accuracy* of the back-end knowledgebase, which intern is regulated by information relevancy and redundancy issues. Existing work on activity discovery have not made any attempt to ensure *relevancy* and *non-redundancy* of discovered knowledge (i.e., location-specific activities). Moreover, majority of these work have utilized body-worn sensors, images or human GPS traces and discovered generalized activities that do not convey any location-specific knowledge.

In this thesis, we address the mentioned issues with serious concern and propose an effective solution to discover Location-specific Activity Network, in short LANet from location-aware reviews. The information network LANet serves as an accurate, enriched and unified knowledgebase of a Location-aware Activity Recommendation System. While building LANet, we also introduce novel ideas like, *activity-based location similarity detection* and *measuring uniqueness, generality / speciality of an activity at a particular location* to enrich the said knowledge base to a great extent. Experimental results show the information richness and accuracy of the proposed knowledge base which is comparable to human perception and accounts for our success in achieving the desired solution.

# DEDICATION AND ACKNOWLEDGEMENTS

Throughout my life two persons have always been there during those difficult and trying times. They are my parents whose continues support, concern and valuable advices made me what I am now. I dedicate this thesis to them for their continuous encouragement that kindled an inspiration in me to pursue the research.

**Dr. Dhaval Patel** has been the ideal thesis supervisor who guided me like a shadow to pursue quality research throughout my entire graduate research period. He truly understands what hard work and dedication can bring to one's life. I am grateful to him for his sage advice, insightful criticisms, and patient encouragement that aided the writing of this thesis in a quintessential way.

 I would also like to thank **Amit Sharma**, **Abhinna Agarwal** and **Shreyash Srivastava** who were good friends, and were always willing to help and give their best suggestions which helped me to overcome setbacks and stay focused on my thesis work.

# TABLE OF CONTENTS

## INTRODUCTION

*"We wander for distraction, but we travel for fulfillment."*

**-Hilaire Belloc**

MOVEMENT is an integral part of human daily life. Most of the times, people visit various locations to perform activity according to their preference. For example, people visit a restaurant with the purpose of having food, visit a shopping mall to do shopping and so on. And whenever they feel something interesting about those locations, they share their experiences with their friends. These location-aware experiences mostly include what activity they have performend there and how they feel about it. Recent growth in smart-phone enabled location-based social networking (LBSN) platforms like yelp[1], foursquare[2], Brightkite[3] etc. has given them the opportunity to jot down their experiences in terms of location-aware reviews. These reviews acts as a great resource for getting information about location-specific activities. *Foursquare* allows its users to *"check-in"* at venues using a mobile website, text messaging or a device specific application which is selected from a list of venues nearby located by the application [15]. As of 2008, another popular LBSN, *Yelp.com*, has listed 4,000 reviews for restaurants in San Francisco and has been active in 18 other US based metro areas including Boston, Chicago, New York, Washington, D.C., San Diego and Los Angeles [12]. As of June, 2013, *Yelp* has been populated by over 42 million reviews for various locations in US [30]. These reviews contain information about users' personal experiences for location along with the category information of the location. Since, the popularity of LSBNs and demand in smartphone is growing; a large amount of location-specific community-authored reviews is available for analysis.

---

[1]http://www.yelp.com/
[2]https://foursquare.com/
[3]http://brightkite.com/

## 1.1   Location-aware Activity Recommendation System - LActRS

The location-specific community-authored reviews act as a potential resource to develop a novel Location-aware Activity Recommendation System (LActRS) as shown in the Figure 1.1. Such system analyzes location-aware reviews and extracts location-specific popular activities. The extracted activities are maintained in the form of a back-end knowledgebase and leveraged to support the location-aware activity recommendation applications.

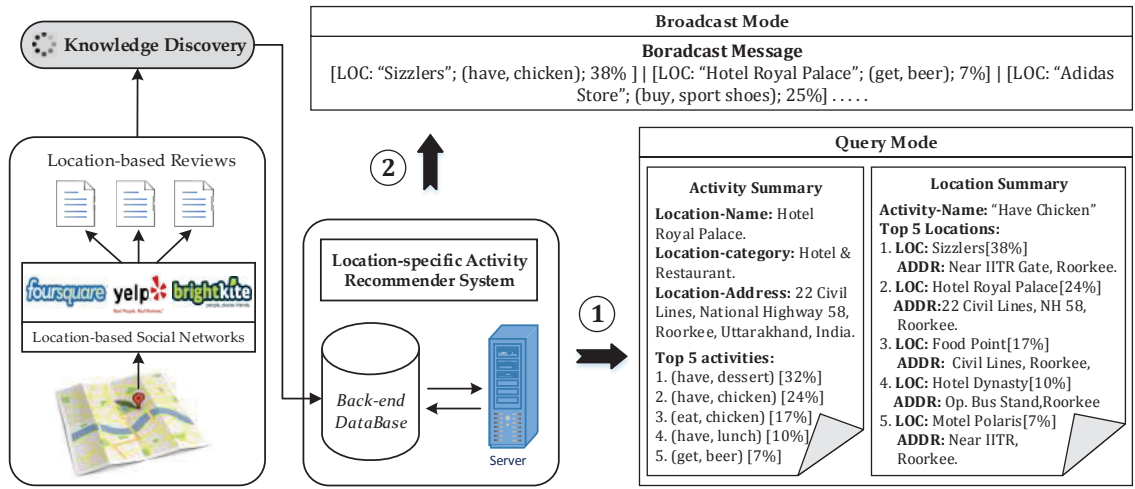

Figure 1.1: Working of Location-aware Activity Recommendation System.

In particular, LActRS can operate in two modes as follows:

1. **Query mode.** In Query mode, user issues location-specific activity query and waits for the LActRS to respond. Here, location-specific activity query either involves identifying activities performed at a particular location or discovering a set of nearby locations that support a given activity. For example, the right bottom rectangle in Figure 1.1 displays top-5 popular activities at location "*Hotel Royal*

*Palace*" situated in Roorkee, India and top-5 locations that support the activity "*have chicken*" in Roorkee. Here, the popularity of an activity at a particular location is considered based on the percentage of users who talked about that activity while reviewing the location.

2. **Broadcast mode.** In Broadcast mode, system `LActRS` periodically broadcast the popular activities supported by a set of nearby locations. Here, the end users tune their electronic gadgets to the nearest broadcast server, receive the localized broadcast messages, and choose the desired information according to their needs.

In summary, `LActRS` makes decision in query mode, whereas user makes decision in broadcast mode.

The successful development of `LActRS` can help people to deal with various decision making problems in automated way. For example, think about a situation- *a tourist is looking for a nearby restaurant in a new place*. To resolve this kind of problem, the tourist generally asks local people for suggestions. But the suggestions given by the local people are generally biased, incomplete, diverse and depend on their personal choices. Thus, sometimes it is very difficult for a tourist to choose the right answer from them. Further, conversational language is also an issue to be considered. In this situation, the Query mode of `LActRS` can help the tourist to figure out the desired location without wasting time in asking people and making decision from their diverse suggestions. Also, sometimes if he/she needs to compare a set of nearby locations based on the activities, he/she can tune his mobile to operate in broadcast mode and analyze the received broadcast information using on-the-fly skyline operations [26].

## 1.2 Relevancy and Redundancy in Activity Recommendation

One of the key factors behind successful implementation of `LActRS` depends on how accurate its knowledge base is. In our case, the knowledge base contains a set of location-specific activities. We can utilize an approach suggested by Dearman et. al. in [11] to infer the location-specific activities from community-authored reviews. However, this approach does not consider two important issues such as *relevancy* and *non-redundancy* of the discovered location-specific activities (see Chapter 2). For example, "*watch cricket match*" is mostly an irrelevant activity for location like "*restaurant*". Similarly, "*had chicken*" and "*eat chicken*" represent redundant activities. The presence of irrelevant and redundant information in the knowledgebase has adverse effects on the performance of any recommendation system. For example, separately, "*have chicken*" and "*eat chicken*" activities may not be the most frequent activities performed at the location "*Hotel Royal Palace*" as we can see in the Figure 1.1. But, if we merge these two redundant activities together, the resultant activity can occupy the top most position in the top-k activity list based on its popularity. This example suggests that *irrelevant* and *redundant* information can cause incorrect ranking of results and hence, mislead the user. Moreover, the presence of irrelevant and redundant information may limit the amount of useful information that can be pushed from `LActRS` to the client in broadcast environment [16].

## 1.3   Some Novel Problems . . .

Apart from *relevancy* and *non-redundancy* issues of the location-specific activities, some location-specific activity related problems still remain unaddressed. For example, inferring spatial importance of activity which can enrich the knowledgebase further by broadening the range of possible activity-related queries. The spatial importance of an activity can be measured by identifying the *uniqueness of that activity with respect to the neighbourhood locations*, which basically shows the region within which it is only performed at a particular location and nowhere else. Using the knowledge of activity uniqueness, user can decide how much he/she needs to travel to perform an activity if he/she has skipped a location that is well-known for the same. Thus, it is essential to incorporate the spatial importance of an activity into the knowledgebase of `LActRS`.

Moreover, none of the existing works in the area of location-specific activity recommendation has addressed the problem of finding similarity between locations based on the common activities supported by them. Such knowledge of activity-based location similarity can be leveraged to recommend alternate locations to the user and the set of common activities that can be performed there.

In addition, the existing works on location-specific activity recommendations do not say anything about how their proposed system can be used for location-based skyline query processing [26] purpose and its implementation in wireless broadcast environment [16]. This kind of spatial applications are very helpful for users in emergency situations when it becomes very difficult for the user to query explicitly to the recommender system.

Thus, considering all these novel problems and different solutions for each of them, if such a knowledgebase can be built that can support a wide range of location-specific activity related queries, it would be beneficial for users to use the knowledgebase and also helpful in developing spatial applications with an extended set of features.

## 1.4 Thesis Contributions

Need for an accurate, non-redundant and enriched knowledgebase is an important concern in developing location-aware activity recommendation system (LActRS). Because, the performance of such recommendation system depends on the *relevancy*, *non-redundancy* and *accuracy* of the location-specific activities that take part in the formation of its back-end knowledgebase. In this thesis, we investigate the problem of discovering *relevant* and *non-redundant* location-specific activities from location-aware reviews and use these inferred activities to build `Location-aware Activity Network` (LANet), in short `LANet` which can serve as an accurate, enriched and unified back-end knowledgebase for the said location-aware activity recommendation system.

`LANet` is basically an information network modeled in terms of a property graph `G=` `<V,E>` where `V` is the set of location and activity nodes and `E` is the set of links that connect these nodes to unite the network. The links are of two types: **(1)** *link that connects two locations and represents similarity in between them* and **(2)** *link that relates an activity with a location*. Each link and node is associated with a set of well-defined properties that helps in storing the knowledge in this network. To the best of our knowledge, no other work has ever discussed about building structure similar to `LANet`.

In order to build `LANet` from location-aware reviews, we introduce the `LANet` discovery process which concerns about the *relevancy* and non-redundancy issues associated with the activities as well as provides an integrated solution to the untouched problems stated in the previous section. The mentioned process takes community-authored location-specific reviews as an initial input, extracts relevant and non-redundant location-specific activities and finally builds the information network-`LANet`.

The major research contributions in our effort to discover `LANet` in this thesis are summarized in the following points:

- We address the issues of *relevancy*, *non-redundancy* of location-specific Activities and introduce novel idea like building Location-specific Activity Network from the community-authored location-aware reviews. The proposed `LANet` not only portrays the spatial relationships among locations and activities but also serve as an accurate, enriched and unified knowledge base for any location-aware activity recommender system.

- We propose a novel technique to discover *relevant* and *non-redundant* location-specific activities from location-aware reviews, called `ActMiner` which is the core processing module of `LANet` that aids in forming activity nodes. We compare `ActMiner` with an existing location-specific activity extraction approach [11]. Experimental results suggest that our method extracts more accurate and meaningful activities from location-aware reviews compared to those inferred by the existing approach.

- We introduce novel ideas and techniques for detecting activity-based location similarity and computing *Uniqueness* value for an activity at a given location and enriched `LANet` with these knowledge. Such ideas are helpful in resolving problems like *alternate location finding* and *inferring spatial importance of an activity at a given location* (see section 1.3). Experimental results shows that our proposed techniques successfully resolve these novel issues and discover knowledge that conforms to the facts of the real-world.

## 1.5   Organization

The rest of this thesis is organized as follows. Chapter 2 discusses about the related works done in the area of activity recognition by following various approaches. We have categorized the related work into three major categories based on their approaches and discussed their limitations in discovering location-specific activities with examples and statistics.

Chapter 3 presents a summarized view of our proposed solution. In this chapter, we first introduce preliminary ideas that helps in understanding the proposed solution followed by an overview of the proposed `lANet` discovery process where we show how we utilize the location-aware reviews to discover `LANet` in successive stages.

Chapter 4 introduces the details of `ActMiner`- the activity discovery module of `LANet` discovery process. In this chapter, we describe the working of `ActMiner`, i.e, how it discovers location-specific *relevant* and *non-redundant* activities from location-aware reviews by following a novel three-phase Discover-Filter-Merge based approach.

Chapter 5 describes the details about how `LANet` is formed in successive stages from location-aware reviews by following novel techniques and utilizing activities inferred by `ActMiner`.

Chapter 6 presents the experimental evaluation of our proposed `LANet` discovery process, which shows that we have successfully achieved our desired objective with considerable accuracy. Finally, chapter 7 presents the concluding remarks about the thesis and provides suggestions for future research.

HUMAN activity recognition is an active area of research due to notable advancement in sensor and GPS technologies. A significant amount of research works have been done in past decade in this area and myriads of useful applications have been developed. Whereas most of these existing works are mostly dedicated to identifying physical activities performed by a human being [17][22][23][18][9][24][8][27] and the activities supported by a location [14][19][32][31][28], some of the works in recent years have paid attention to predict location-based activity that a user is possibly going to perform next by mining user's location history [29]. A careful analysis of the wide spectrum of research done in this area classifies the existing works into three major categories as follows:

## 2.1   Activity Recognition using Sensors and Image Processing Techniques

The first category recognizes human activities by analyzing human body movement and gesture [17][22][23][18][9][24][8][27]. Most of these techniques have employed low-fidelity sensors like RFID tags, motion detectors, accelerometers to monitor human (mostly pedestrians) and then discover human body-movement related activities [22], household activities [23], activities associated with transportation mode [18] etc. A brief description of the mentioned related works done in this area can be found in the paper [17]. In case of activity recognition using image processing techniques, work [9] presents the design and implementation of an activity recognition system for wide area aerial video surveillance using Entity Relationship Models (ERM). Here, by incorporating reference imagery and Geographical Information System (GIS) data, the work has shown a procedure to track moving objects and thus, recognize activities by analyzing their traffic patterns. Other notable works done in this area are developing a system for human activity recognition in video sequences [24], designing "smart video" system to track pedestrians and detect suspicious motion or activities at or near critical transportation assets [8], human activity recognition based on R transform [27] etc.

**Limitation.** Majority of these techniques recognize only human physical activities (like walking, running, eating, drinking, sitting etc.), but not location-specific activities. These physical activities are basically activities of daily living (ADLs) [13] which are are very generalized in sense and performed irrespective of location. Hence, they do not give any meaningful or distinguishable information about a particular location.

## 2.2   Inferring Activities from GPS Traces

The second category discovers human activities by analyzing GPS trajectories [14][19] [32][31][28]. As GPS trajectories capture human movement over various geographical locations, the information about user's location history is being utilized by researcher to infer user's activity. For example, paper [14] proposes an algorithm that automatically annotates raw GPS trajectories with the activities performed by the users at various places. Next, it considers the stops points of the user to infer Point of Interests (POIs) and based on the category of the POI and a gravity law based probability measure, it infers the activity performed at various locations. Apart from inferring user activities, GPS trajectories are also used to discover the interesting locations visited by many people. For example, paper [19] uses hierarchically structured conditional random fields to generate a consistent model of a person's activities and places. Paper [32] determines location features and activity-activity correlations from the geographical databases and the web to gather additional inputs and applies a collective matrix factorization method to mine interesting locations and activities from user's location and activity histories. Other notable works in this domain are recognizing human activity with trajectory data in multi-floor indoor environment [31] and joining large set of trajectories with activities using duplication reuse techniques [28] etc.

**Limitation.** We noted that the activities inferred using these approaches are general activities such as "*eating*", "*shopping*" etc. This is because, using GPS information we can only detect the category and semantics of the given location and using this, we can infer possible activities which is very generalized in sense. For example, if the category of the location is "*restaurant*", the possible activity will be "*eating*", but we can't tell specifically whether that place is well-known for *eating chicken* or *eating some vegetarian food item*. So, these methods are unable to detect specific activities for which the place may be well-known for.

## 2.3   Inferring Activities by mining Location-specific Reviews

The last category analyzes user generated contents that are obtained from location-aware online social media. The location reviews play an important role in location-specific activity discovery. Because, they are written mostly by people who have real-life experiences of visiting those places and have performed certain activities there. By mining these reviews, we can infer location-specific activities, which are not possible using GPS traces.

The only paper that concerns about inferring human activities by mining location-specific reviews is paper [11] which we have considered as the base paper. They use sentence tokenizer to parse the review text into its individual sentences and then employ part-of-speech tagger to identify the verbs and nouns from each sentence. Next, verb-noun pairs are discovered such that noun is located within 5 words following the verb. The discovered verb-noun pairs are represented in their base form and declared as the potential activities supported by the location. Using 14 test locations, the approach has shown that the majority of the 40 most common results per location (determined by verb-noun pair frequency) are actual activities supported by the respective locations which have achieved a mean precision of up to 79.3%.

Although the procedure is very simple, it is accompanied with some major limitations-

- **Limitation in Activity Discovery.** The baseline approach has only considered the "*nearest*" nouns within 5 words following the verb to generate the (verb, noun) and (verb, noun phrase) pairs that represent location-specific activities. But, the approach hasn't said anything about how the pairs will be generated if the noun occurs before verb which is mostly seen when the sentence is in passive voice.

For example, in the sentence- *"The food was served by the waiter at around 9 pm.",* the baseline approach will infer (*served, waiter*) as the activity pair as noun "*waiter*" appears within 5 words vicinity following the verb "*served*". But, the meaningful activity that this sentence tells about is (*served, food*) which the approach has failed to detect. The reason behind this failure is that the approach has only used the "*distance*" between a verb and a noun for pair generation without considering the "*dependency between the words*" which is the key factor for ensuring meaningfulness of the discovered activities.

- **Limitation in Recommendation.** The baseline approach has concluded that majority of the 40 most common results per location are the *actual activities supported by the respective locations*. So, if the user wants to know whether a specified activity (which doesn't come under top 40) can be performed at a particular location or not, the approach will not be able to give answer with a certain accuracy.

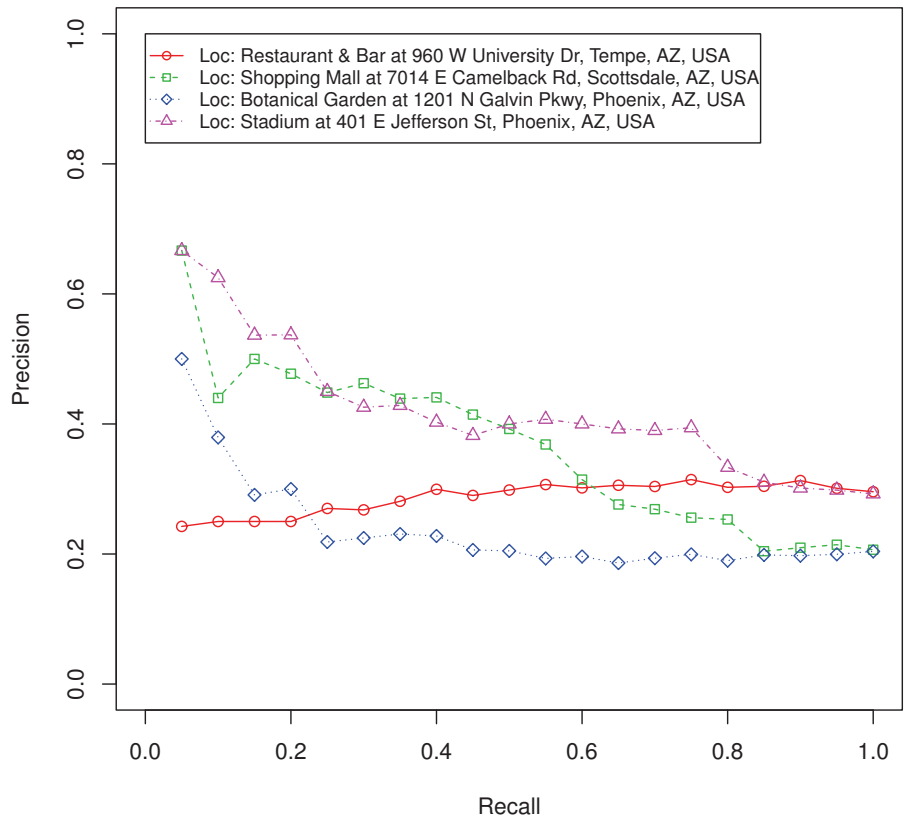

Figure 2.1: Recall vs. Precision graph.

- **Issues with Relevancy.** The baseline approach has not considered the issues with relevancy of the inferred activities (as discussed in section 1.2). So, the top-40 list of activities recommended by the approach may contain some irrelevant activities whereas some relevant activities have been thrown out from the list due to their low frequency of occurrence compared to the irrelevant ones.

We have implemented the baseline approach and tested the relevancy of the concept ("*noun*" part of the activity) associated with the top 500 inferred activities using ConceptNet [22] for each of 4 locations (having more than 200 reviews) selected from the yelp data set. Figure 2.1 shows the result plotted in the form of recall vs. precision graph. Here, the baseline approach has only achieved a maximum precision up to 66.7% on a recall of 5% (in case of 4th location) and a minimum precision of 18.98% on a recall of 80% (in case of 3rd location). So, we can conclude that on an average, only 40% of the top-k activities recommended by the baseline approach are relevant with respect to the category of the location.

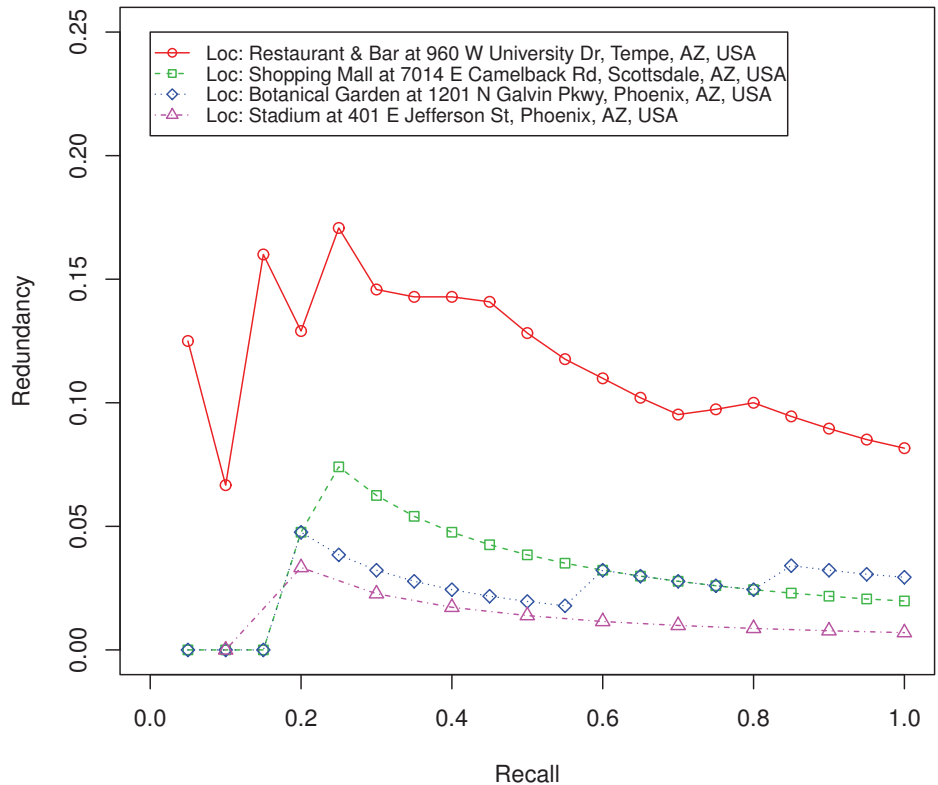

Figure 2.2: Recall vs. Redundancy graph.

- **Issues with Redundancy.** The baseline approach has also not taken any step to resolve the issues involved with redundant activities. We have performed experiments to evaluate the percentage of redundancy present within a set of relevant activities obtained from the top 500 activities for the same set of 4 locations as stated in the previous point. We plotted the experimental results in terms of a recall vs. redundancy graph as shown in the Figure 2.2. Here, the $1^{st}$ location leads among the four with the maximum redundancy of 17% found on a 25% recall within its relevant activity set. The $2^{nd}$ location posses maximum redundancy of 7.4% on a recall of 25% and so on. This proves that the baseline approach has also not taken any step to eliminate redundancies present in its recommended activity set.

All these limitations give us reasons to rethink about providing effective solutions of the stated problems. In summary, we have studied the requirements for designing an efficient location-aware activity recommendation system and observed that, none of the existing approaches has met these requirements. This accounts for our effort to build the information network `LANet` to achieve the desired solution.

## OVERVIEW OF THE LANet DISCOVERY PROCESS

BEFORE introducing the details of the proposed `LANet` Discovery process as discussed in Chapter 1, in this chapter we present a high level overview of the said process. The chapter is divided into two sections. In the first section, we have primarily discussed about some preliminary ideas required to formulate the solution. This also includes the type of data we have worked with to build the proposed solution- `LANet` and a formal problem statement stating clearly the problem we have solved in this dissertation. The next section, presents a glimpse of the `LANet` Discovery process, where we have discussed about various processing done to discover `LANet` in a summarised way. In particular, we have shown how the proposed knowledgebase for activity recommendation evolve in successive processing stages to gain its final shape.

## 3.1 Preliminaries

Let $L = \{L_1, L_2, L_3, ..., L_m\}$ be the set of $m$ locations and $cat_i$ be the list of categories for location $L_i \in L$. In our database, each location $L_i$ is associated with a set of reviews $Rset_i = \{R_i^1, R_i^2, R_i^3, ..., R_i^n\}$, where $R_i^j$ denotes the $j^{th}$ review of location $L_i$. The review $R_i^j$ is written in the form of textual description. Table 3.1 shows a snapshot of the category and reviews for location $L_i$.

**Definition 3.1. Activity.** An activity performed at a particular location is defined as the combination of a verb with a noun or noun phrase. An activity is meaningful if it represents some "*doing*" sense. It is represented as $A_i^j = (verb, noun/noun\ phrase)$ and stands for the $j^{th}$ activity performed at Location $L_i$. Here, a *noun phrase* is a collocation of nouns with which a verb forms association.

Table 3.1: A Snapshot of $(cat_i, Rset_i)$ for a given Location $L_i$

| $cat_i$ | {"*Restaurant*","*Hotel*" } |
|---|---|
| **Review** | **Text** |
| $R_i^1$ | "*Car driving and sometimes, trying out good new foods are my hobbies. Yesterday I came here for dinner. The food was served late and I had to wait for a long time. But, I enjoyed chicken tikka masala and had a great time there.*" |
| $R_i^2$ | "*Had dinner with my old friends. The food was awesome... I liked butter nun very much.*" |
| $R_i^3$ | "*Yesterday, we celebrated my aunt's birthday in the banquet hall. Everyone enjoyed a lot...*" |

For example, the combination of verb "*celebrated*" and noun "*birthday*" in Review $R_i^1$ denotes an activity performed at location $L_i$. Similarly, (*had, dinner*) and (*liked, chicken tikka masala*) represent activities in review $R_i^2$. The popularity of an activity $A_i^j$ performed at a given Location $L_i$ is measured by "*Activity Frequency*" as given below.

**Definition 3.2. Activity Frequency.** The frequency of an activity $A_i^j$, denoted as $AF(A_i^j)$, is the number of distinct reviews that has mentioned about $A_i^j$ at location $L_i$.

For example, if the activity $A_i^j$ = (*eat, chicken*) appears in 5 distinct reviews in the review set $Rset_i$ of location $L_i$, $AF(A_i^j)$ is 5. The high $AF$ value of an activity suggests that many people have written those reviews and have talk about this activity.

**Definition 3.3. Concept.** The concept associated with an activity $A_i^j$ is defined as the *noun* or *noun phrase* part of an activity.

For example, in the activity (*had, dinner*), the noun part "*dinner*" is the concept associated with that activity. Similarly, "*chicken tikka masala*" is the concept associated with the activity (*liked, chicken tikka masala*).

**PROBLEM STATEMENT.** Given the categories and review sets for all $m$ locations, i.e. $\{(Cat_i, Rset_i) \mid 1 \leq i \leq m\}$, our proposed solution discovers set of *relevant* and *non-redundant* location-specific activities $Aset_i = \{A_i^j \mid 1 \leq j \leq r\}$ for each location $L_i$, where $A_i^j$ is an activity performed at location $L_i$ and builds the information network- `LANet`.

## 3.2    A Glimpse of the LANet Discovery Process.

In this thesis, we propose *LANet Discovery Process* as an efficient solution of the stated problem statement. The objective of this process is to discover location-specific activities from community-authored location-aware reviews and form a unified and enriched knowledgebase of locations and their supported activities represented in the form of a graph data model. The overall process of `LANet` discovery be broadly devided into two primary processing task, viz. **(1)** *discovering activities from location-aware reviews* and **(2)** *formation of the knowledgebase by utilizng those discovered activities*. For the activity discovery task, we introduce a novel technique called `ActMiner` (see chapter 4) which is capable of inferring relevant and non-redundant activities from location-aware reviews with a considerable accuracy. We utilize these inferred activities and the locations where these activities are performed in the *graph formation task* in a systematic way to build the desired information network- `LANet` (see chapter 5).

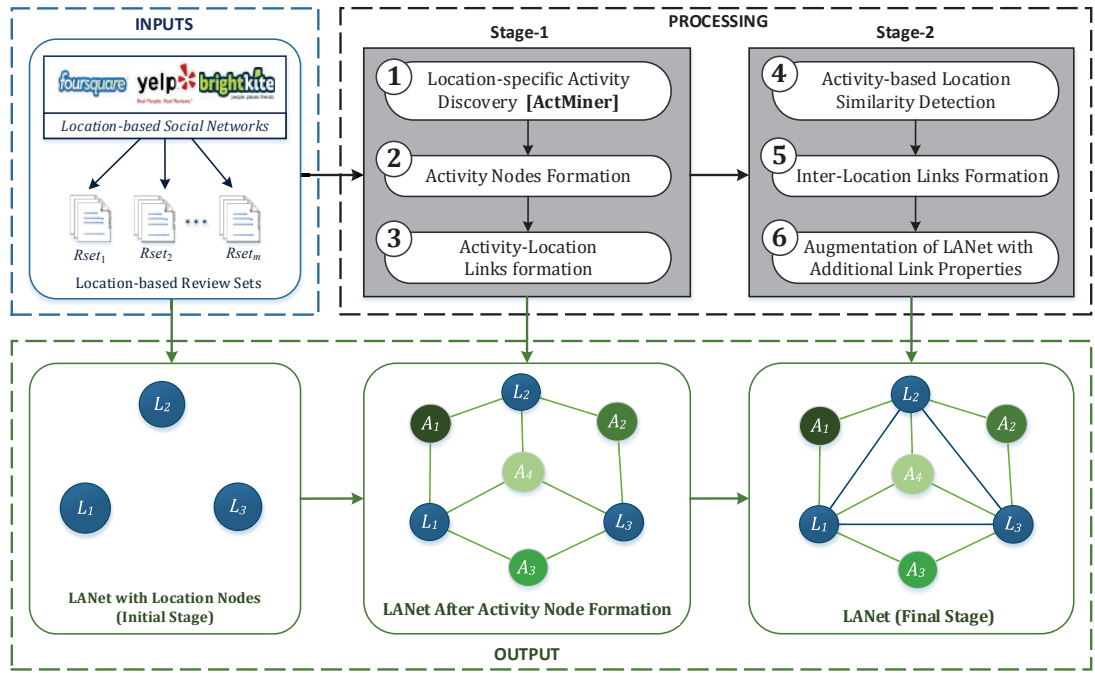

Figure 3.1: Evolution of LANet through Successive Stages.

*LANet Discovery Process* is a sequential graph development process where the information network evolves in successive stages to gain its final shape. Figure 3.1 shows a high level overview of `LANet` discovery process with evolution of `LANet` in each stage.

Initially, `LANet` contains $m$ location nodes- $L_1, L_2, L_3, ..., L_m$ for which there exist review sets. In Stage-1, three processing tasks are performed. First, activities are inferred from the location-aware reviews using `ActMiner` (step-1). Next, activity nodes are formed using those inferred activities (step-2). Then, we connect activity nodes with their supporting location nodes using *Activity-Location Links* (step-3). While, formation of *Activity-Location Links*, we associate each link with a set of property values containing information about the relations between the location and their specific activities. At this moment, processing of stage-1 ends and a LANet is formed with a set of activity nodes associated with their supporting locations.

Next, stage-2 starts. In this stage, first similarity between locations are discovered based on their supported activities (step-4). Here, two locations are considered as similar if they have many common activities in between them. Such similar location pairs are then connected by *Inter-Location Links* enriched with a set of properties containing information about their similarity and common activities (step-5). Finally, we discover additional properties like activity popularity index (specifies relative popularity of an activity at a given location compared to the other locations where the activity is performed) and Boundary of uniqueness (specifies uniqueness of an activity at a given location compared to the other locations where the activity is performed) and augment *Activity-Location Links* with these property values for the further enrichment of `LANet` (step-6). In this way, stage-2 ends and we obtain the desired information network-`LANet` ready to serve as an enriched knowledgebase of a location-aware activity recommendation system.

Algorithm 1 briefly presents the high level overview of the processing done to discover `LANet`. Input to our algorithm is a set of location specific reviews. In Lines 1-3, we initialize a `LANet` with a set of location nodes $L_1, L_2, L_3, ..., L_m$. The remaining process of Algorithm 1 is divided into two stages namely, *Activity Nodes and Activity-Location Links Formation Stage* (Lines 4-20) and *Inter-Location Links and Additional Property Augmentation Stage* (Lines 21-29). In *Activity Nodes and Activity-Location Links Formation Stage*, relevant and distinct activities are extracted for each location using `ActMiner` (Lines 4-16) and then, activity nodes are created for each of those discovered activities along with *Activity-Location Links* (Lines 17-20). In *Inter-Location Links and Additional Property Augmentation Stage*, we link location nodes based on the similarity between activities that are being performed at that locations (Lines 21-25), compute *activity popularity index* and *uniqueness* value of the location-specific activities and augment *Activity-Location Links* with these property values (Line 26-28). Chapters 4 and 5 discuss details of both stages.

---

**ALGORITHM 1: Discover LANet**

---

**Data**: $\{(Cat_i, Rset_i) \mid 1 \le i \le m\}$: list of categories and review sets for $m$ number of locations.

**Result**: LANet- the information network of locations and activities.

    /* Initialization Stage                                         */

**1**   **for** *each Location $L_i$ in $L$* **do**

**2**      **Create** a node for $L_i$ with location node properties in LANet;

**3**   **end**

    /* Activity Nodes and Activity-Location Links Formation Stage    */

**4**   **for** *each Location $L_i$ in $L$* **do**

**5**      **for** *each Review $R_i^j$ in $Rset_i$* **do**

**6**          **Generate** Activity set $Aset_i$ for $L_i$ by extracting activities from $R_i^j$;

**7**      **end**

**8**      **for** *each Activity $A_i^k$ in $Aset_i$* **do**

**9**          **if** *$A_i^k$ is irrelevant with respect to $cat_i$* **then**

**10**              **Remove** $A_i^k$ from $Aset_i$;

**11**          **end**

**12**      **end**

**13**      **for** *each Activity $A_i^k$ in $Aset_i$* **do**

**14**          **Merge** $A_i^j$ with $A_i^k$ for some $A_i^j \in Aset_i$ , if $A_i^j$ and $A_i^k$ are redundant to each-other;

             **Replace** $A_i^k$ with the merged activity and remove $A_i^j$ from $Aset_i$;

**15**      **end**

**16**   **end**

**17**   **Generate** global activity set $Aset$ such that $Aset = \bigcup Aset_i$.

**18**   **for** *each Activity $A_i^k$ in $Aset$* **do**

**19**      **Create** a Node for $A_i^k$ and link it with $L_i$ using "*Is_Performed_At*" relation along with
       relationship properties;

**20**   **end**

    /* Inter-Location Links and Additional Property Augmentation Stage    */

**21**   **for** *each Location $L_i$ in $L$* **do**

**22**      **Create** $sim\_set_i$ for $L_i$, where $sim\_set_i = \{L_j \mid L_j \in L$ and similarity$(L_i,L_j)>0\}$;

**23**      **for** *each Location $L_i$ in $sim\_set_i$* **do**

**24**          **Link** node $L_i$ with node $L_j$ using "*Is_Similar_To*" relation and add corresponding Link
           properties;

**25**      **end**

**26**      **for** *each Activity $A_i^k$ in $Aset_i$* **do**

**27**          **Compute** Activity popularity index and Boundary of uniqueness value of $A_i^k$ and
           augment Link $(A_i^k, L_i)$ with this property value;

**28**      **end**

**29**   **end**

---

## ACTMINER: DISCOVERING ACTIVITIES FROM LOCATION-AWARE REVIEWS

K NOWLEDGE of Location-specific activities is the heart of `LANet`. In this chapter, we discuss the *core processing module of* `LANet` *discovery process* known as `ActMiner` proposed in [21]. Given the categories and review sets for all $m$ locations, `ActMiner` generate activities for each location using Discover-Filter-Merge technique (See Figure 4.1).

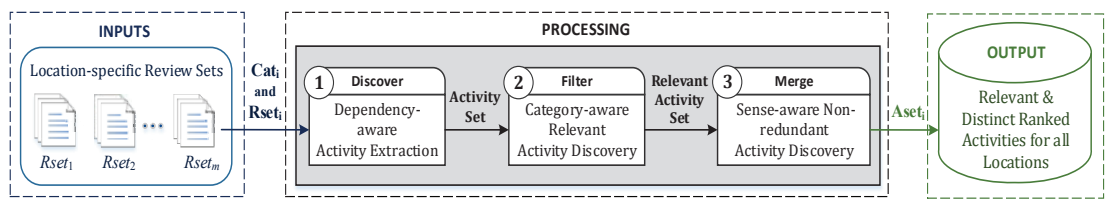

Figure 4.1: Working of ActMiner.

The Discover-Filter-Merge technique process review set $Rset_i$ for location $L_i$ in three sequential phases. In the *first phase*, `ActMiner` systematically uses natural language processing techniques to discover potential (verb, noun) and (verb, noun phrase) pairs that represent meaningful activities. In the *second phase*, *relevant* activities are discovered from the output activities of phase I, using ConceptNet and category information of the location and irrelevant activities are filtered out. In the *last phase*, the proposed solution identifies the redundant activities and merges them together into a single activity. At the end, the discovered *relevant* and *non-redundant* location-specific activities are stored in
a repository along with their frequencies as the final output. In the remaining sections,
we explain details of each phase using $Rset_i$ for location $L_i$.

## 4.1 Dependency-aware Activity Extraction

We can process each review $R \in Rset_i$ and discover any (*verb, noun*) or (*verb, noun
phrase*) present in the review message. However, this simple approach generates a large
number of spurious activities if the relationship between verb and noun is not taken
into account. Thus, we develop a novel dependency-aware activity extraction technique
that utilizes the typed dependency and proximity information between verb and noun to
discover the activity. In particular, we employ a series of NLP operations on each review
$R$ in $Rset_i$, as follows:

1. We use OpennlpSentenceDetector [1] to extract individual sentences from review
   $R$.

2. Next, we tag each word of a sentence with their corresponding part-of-speech
   using Stanford Part-Of-Speech (POS) Tagger [4] and extract verbs, nouns and noun
   phrases (i.e., collocation of nouns) from the tagged sentence. While extracting verbs,
   we have neglected "*was*", "*were*", "*is*", "*am*" etc. verbs and the *auxiliary verbs* like
   "has", "have", "had", "been" etc. that frequently appear in the sentence but don't
   contribute in meaningful activity extraction. For example, in the sentence- "He
   has played the game", "has" is an auxiliary verb of the main verb "played" and
   hence, is neglected during verb extraction. Thus, we get a potential verb set and a
   noun/noun phrase set. The verb set is called *potential* because all the verbs in this
   set have potential to take part in meaningful activity extraction.

3. Simultaneously, we parse each sentence using Stanford Typed dependency parser
   [5] and detect all activities in terms of (*verb, noun*) pairs such that the verb in that
   pair belongs to the potential verb set and the pair has any of the following gram-
   matical relations between the verb and the noun- *dobj*("*direct object*"), *nsubjpass*
   ("*passive nominal object*"), *ccomp* ("*causal complement*" of a verb) or *prepositional
   grammatical relations* (like *prep_to*, *prep_at*, *prep_for* etc.) Although dependency
   parser can report around 51 dependencies [10], we have observed that the afore-
   mentioned four dependencies help us to capture most the meaningful activities.

Table 4.1: Cases where Dependency parsing helps in direct activity extraction

| Types of Sentence | Example | Dependency Relations Used | Relations Inferring Activities |
|---|---|---|---|
| Active Voice | *"I watched the movie yesterday with my friends."* | dobj, ccomp and prepositional relations like prep_to, prep_at, prep_for etc. | *dobj (watched, movie)* dependency infers the activity *(watched, movie)* |
| Passive Voice | *"The match was played between India and Australia."* | nsubjpass | *nsubjpass (played, match)* dependency infers the activity *(played, match)* |

Table 4.1 shows the examples where these dependencies helps to infer meaningful activities present in the sentence.

4. Dependency parser doesn't generate an activity in the form of (*verb, noun phrase*) pair. To address this issue, we utilize the extracted noun/noun phrase set (discussed in step 2) and replace noun part of the detected (verb, noun) pair with the corresponding noun phrase part.

We explain the above mentioned procedure with the help of a sample review $R$ as shown in the Figure 4.2. First step extracts four sentences from $R$. In the next step, 8 nouns, 1 noun phrase and 8 verbs are extracted from the four sentences and a potential verb set and a noun/noun phrase set are formed. Then with the help of the potential verb set and dependency parsing on each of the four sentences, total six activities are discovered, where (*trying, food*) activity is discovered from sentence-1 using dependency relation "*dobj(trying, foods)*", (*came, dinner*) is discovered from sentence-2 using "*prep_for(came, dinner)*", and so on. Simultaneously, POS Tagger detects *chicken tikka masala* as a noun phrase from sentence-4. Thus, pair (*enjoyed, masala*) is converted into the pair (*enjoyed, chicken tikka masala*).

Often, use of noun phrase instead of noun helps in detecting specialized activities. As we can observe that "*chicken tikka masala*" is a specialized category of the food item "*chicken*". Not only that, it also helps us to resolve some discrepancy that may raise if we consider only noun without giving any attention to the use of noun phrase. For example, if we consider the sentence - "*I visited the food court of the city mall yesterday.*", the meaning of the noun phrase "*food court*" is different from the noun "*food*". So, in this case, dependency parsing can't detect the desired pair alone as it infers the pair as (*visited, food*). But if the knowledge of noun phrase is used here, we can easily discover the correct pair (*visited, food court*) which conveys some meaningful activity.

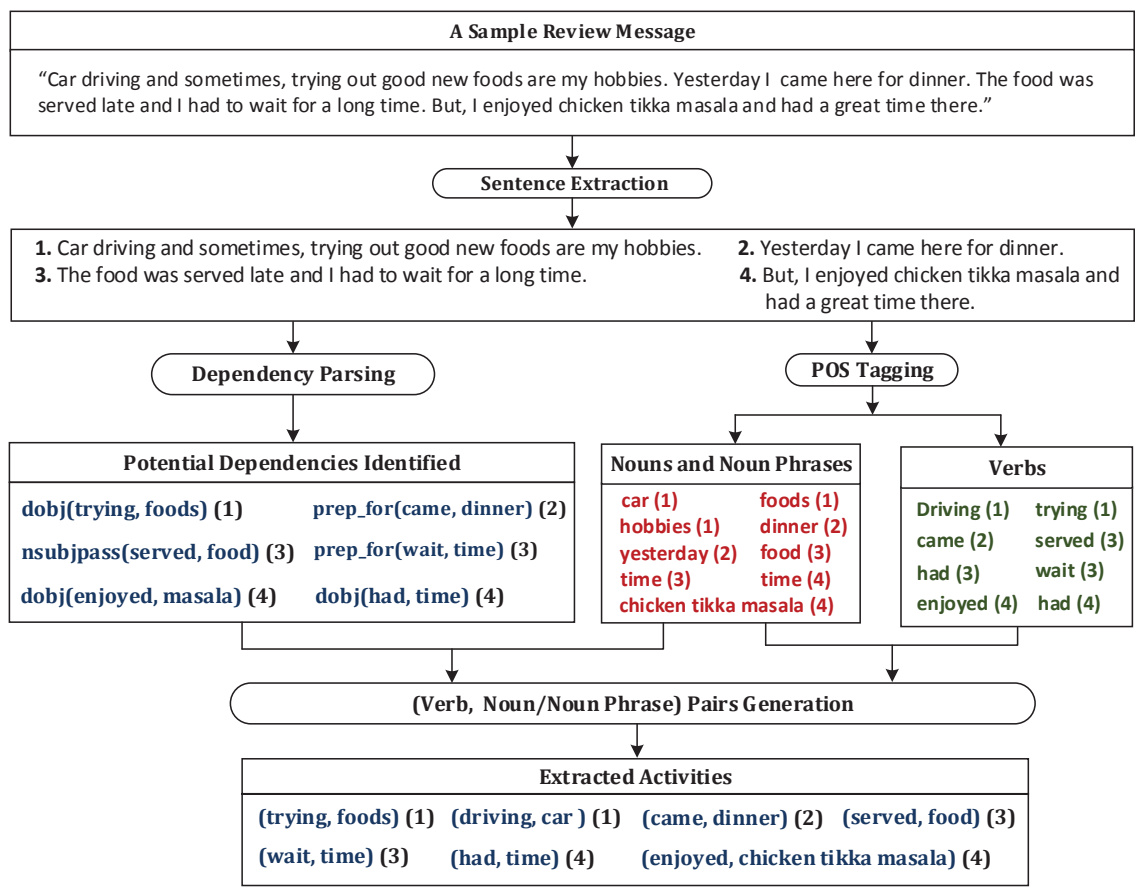

Figure 4.2: An Example of Activity Extraction from Review $R$. The number given in "()" associated with each activity, extracted verb and noun/noun phrase indicates the sentences from which they are extracted.

Sometimes, combination of two nouns, where one noun is derived from a verb in its base form by adding "*ing*" to it, also represent a meaningful activity. In Figure **??**, the word "*driving*" in sentence-1 is actually tagged as noun by pos-tagger, but it is derived from the base verb "*drive*". To deal with this case, while POS tagging a sentence, we have used WordNet [7] to check whether a noun that ends with "*ing*" has a base verb form in its synset or not. If there exists a base verb form for that noun, (for example, "*drive*" is the base verb form present in the synset of noun "*driving*"), we treat such noun as verb and generate pair by associating it with the nearest noun or noun phrase in the sentence.

Thus, (*driving, car*) has been inferred as an activity from sentence-1.

**Post-processing.** Once, activities are discovered from each review $R$ of $Rset_i$, all the verbs and nouns in the (*verb, noun/noun phrase*) pairs are converted into their base forms using WordNet. For example, (*ate, food*) is converted into (*eat, food*). Then, $AF$ values of each of the discovered activities are calculated and discovered activities are stored in $Aset_i$ in the form [(*verb,noun / noun phrase*), $AF$].

## 4.2 Category-aware Relevant Activity Discovery

The activities discovered in previous stage are syntactically correct, but may not represent meaningful activities from the semantic point of view. In particular, there may be activities that cannot be performed at a given location and thus are not relevant. In this paper, we resolve activity relevancy issue by *leveraging the category information of the location*. We use a semantic knowledgebase such as ConceptNet [2] and validate whether the concept associated with an activity (Recall Definition 3.3) conforms to the category of $L_i$ or not. Note that, we relate the conceptual information of an activity with the category of location as the category of a location is conceptual information and the activities are concrete in sense.

The basic observation behind our idea is that the concept of a relevant activity is mostly associated with the category of the location. For example, given the activity (*eat, food*) at location $L_i$, the concept "*food*" is associated with location $L_i$, if category of $L_i$ is "*restaurant*". By exploring ConceptNet, we find that concept "*food*" is related to the concept "*restaurant*" by the relation $\{food \overset{\text{At Location}}{\rightarrow} restaurant\}$. However, in most of the cases, the concept associated with an activity doesn't have any direct relationship with the category of the location. For example, concept "*chicken*" is not directly associated with the concept "*restaurant*" in ConceptNet. However, "*chicken*" is associated with "*food*" and "*food*" is intern related to "*restaurant*". Hence, a simple look up in ConceptNet doesn't give the desired solution for activity relevancy problem. However, considering size of ConceptNet, it is not feasible to explore all possible indirect connections. We present a systematic way of finding out the chain of relations that associates a given concept to the category of a location in ConceptNet.

Given a set of activities $Aset_i$ for location $L_i$, we extract concept by processing each activity in $Aset_i$ and output the *noun or noun phrase* part of an activity as a concept. For each concept, we also maintain a *Concept Frequency*(CF) value which tells about the number of distinct reviews a given concept has been referred. If concept occurs multiple times within a single review, we count it only once. At the end, we get a set of concepts $Cset_i$, where each concept in $Cset_i$ is in the form of $(concept_k, CF_k)$. Figure 4.3 provides example of extracted concepts from a set of activities.

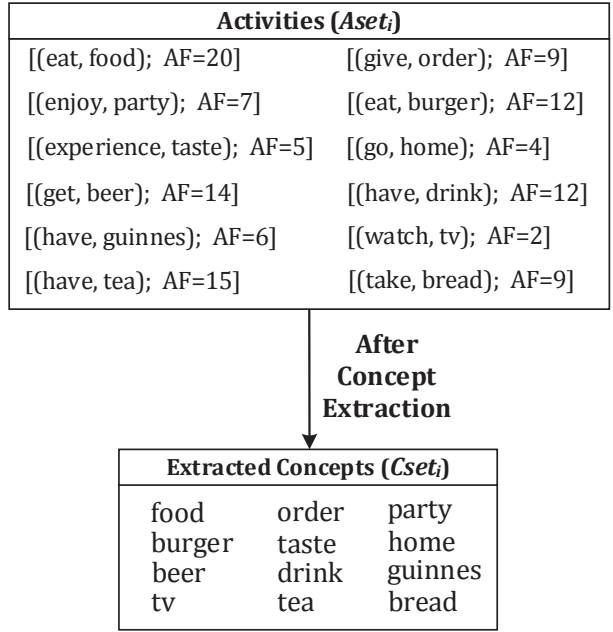

Figure 4.3: Discovering Concepts from Activities.

Let, $Cset_i$ be the set of discovered concepts. Then, we learn a *Category-aware Concept Hierarchy*, denoted as CCH, using the set of extracted concepts $Cset_i$, category list $Cat_i$ and the ConceptNet. The proposed structure organizes concepts coming from $Cset_i$ and $Cat_i$ into a hierarchy such that we can verify whether a concept associated with an activity is related to the category of the location or not. While forming the hierarchy, we use only "*IsA*", "*AtLocation*", "*DerivedFrom*", "*UsedFor*" and "*RelatedTo*" relations of the ConceptNet to learn the CCH. The first three relations "*IsA*", "*AtLocation*" and "*DerivedFrom*" capture generalization-specialization relationship between two concepts. For example, "*novel*" is a specialized concept of "*book*" - is represented by $\{novel \overset{\text{IsA}}{\to} book\}$, "*food*" is found at "*restraurant*" - is represented by $\{food \overset{\text{AtLocation}}{\to} restaurant\}$ and "*shopper*" is derived from "*shop*" - is represented by $\{shopper \overset{\text{DerivedFrom}}{\to} shop\}$. The remaining two relations "*UsedFor*" and "*RelatedTo*" are used for linking related concepts. For example, in the relation- $\{kitchen \overset{\text{UsedFor}}{\to} cook\}$, "*kitchen*" and "*cook*" are related concepts and in $\{cake \overset{\text{RelatedTo}}{\to} birthday\}$, "*cake*" and "*birthday*" are related concepts. Now, we formally define CCH.

**Definition 4.1. Category-aware Concept Hierarchy (CCH).** A **C**ategory-aware
**C**oncept **H**ierarchy for a given location is a tree-based hierarchy and represented by
triplet $< Lv, C, E >$, where $Lv = \{lv_1, lv_2, lv_3, ..., lv_k\}$ is the collection of levels , $C = \{C_1, C_2, C_3, ..., C_k\}$ is a collection of concept sets with $C_i$ be the set of concepts at level
$lv_i$, and $E$ is the set of labeled arcs that connect concepts lying in the same level or in
successive levels. The structure satisfies the following 3 properties:

- $C_1 = Cat_i$ and $C_i \subseteq Cset_i$ where $2 \le i \le k$.

- If any two concepts $c$ and $c$' at the same level are connected by a labeled arc $e_r$,
  then the label of $e_r \in \{\text{"RelatedTo"}, \text{"UsedFor"}\}$.

- If any two concepts $c$ and $c$' at level $lv_i$ and $lv_{(i+1)}$ respectively, are connected by a labeled
  arc $e_r$, then label of $e_r \in \{\text{"IsA"}, \text{"AtLocation"}, \text{"DerivedFrom"}\}$.

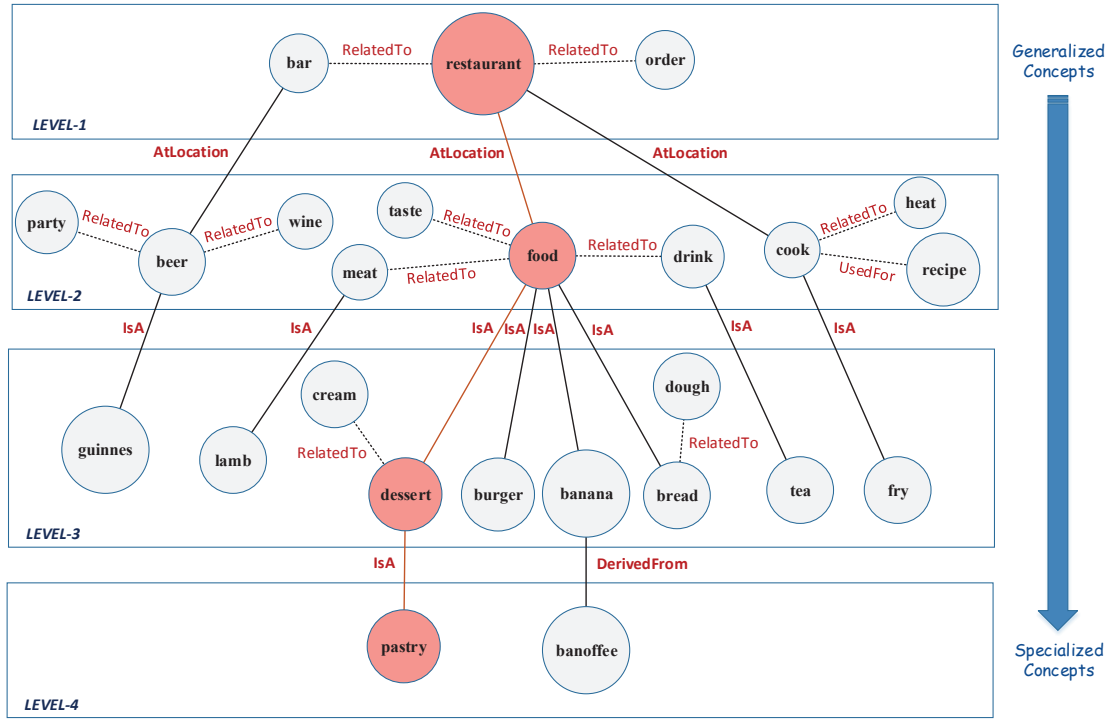

Figure 4.4: An Example of CCH for a location with category "*restaurant*" cum "*bar*".

Figure 4.4 shows an example of CCH for a given location with category "*restaurant*"
cum "*bar*". Such structure not only helps to recognize relevant concepts, but also plays a
major role in categorizing a location-specific activity as generalized or specialized one.

As we can see in the Figure 4.4, each concept that belongs to the CCH is *relevant* with respect to the *category* of the location as it is connected to the *category* of the location by a well-defined chain of relationship. For example, the relevant concept "*pastry*" is related to the concept "*restaurant*" via the following path $\{pastry \stackrel{\text{IsA}}{\rightarrow} dessert \stackrel{\text{IsA}}{\rightarrow} food \stackrel{\text{AtLocation}}{\rightarrow} restaurant\}$. And as we move down the hierarchy, the concepts lying in the lower levels becomes more specialized with respect to the generalized concepts. For example, concepts "*dessert*", "*burger*", "*banana*", "*bread*" are specialized with respect to the concept "*food*".

**CCH Formation**. The First level of CCH is initialized with the concepts from $Cat_i$. Next, we perform two operations *Expand* and *Extend* iteratively to grow the hierarchy in horizontal and vertical dimensions respectively. The *Expand* operation uses "*UsedFor*" and "*RelatedTo*" relations to add concepts from $Cset_i$ into the current level. The *Extend* operation uses "*IsA*", "*AtLocation*" and "*DerivedFrom*" relations to add concepts from $Cset_i$ into the next level. In summary, the *Expand* operation adds concepts that are associated with the concepts lying in the same level and the *Extend* operation adds concepts that are specialized in sense with respect to the concepts lying in just upper level. This iterative procedure stops when the hierarchy cannot grow further. We explain the CCH formation using Example 1.

**Example 1**. Figure 4.5(a) shows the sample inputs for CCH formation process which consist of $Cat_i$ = {"*restaurant*", "*bar*"}, extracted concept set $Cset_i$ containing 15 concepts and the ConceptNet. At first, the CCH is initialized with two concepts "*restaurant*" and "*bar*" and in this way, level-1 is created (See Figure 4.5(b)). Next, the *expand* operation is performed on level-1 and as a result, concept "*order*" from $Cset_i$ is added in level-1 and is linked with "*restaurant*" by relation name "*RelatedTo*". At the same time, concept "*bar*" also gets associated with the concept "*restaurant*" by "*RelatedTo*" relation. Next, the *extend* operation is performed on the expanded level-1 which adds "*beer*" and "*food*" from $Cset_i$ as the child nodes of "*restaurant*", links them by "*AtLocation*" relations and forms level-2. This marks end of Iteration-1 and *iteration-2* starts with execution of *expand* operation on level-2. This operation causes the expansion of level-2 by adding concepts "*party*", "*drink*" and "*taste*" and and linking them with the existing concepts in that level. Once *expand* operation level-2 finishes, *extend* operations starts and the concepts "*guinnes*", "*coffee*", "*tea*", "*burger*", "*pie*" and "*bread*" are added into the CCH by forming level-3 and are linked by "*IsA*" relations with the upper level concepts as shown in the Figure 4.5(b). In this way, *Iteration-2* ends and *Iteration-3* starts. But in *iteration-3*, none of the concepts from $Cset_i$ gets added into the CCH. So, the CCH doesn't grow further and the process terminates.

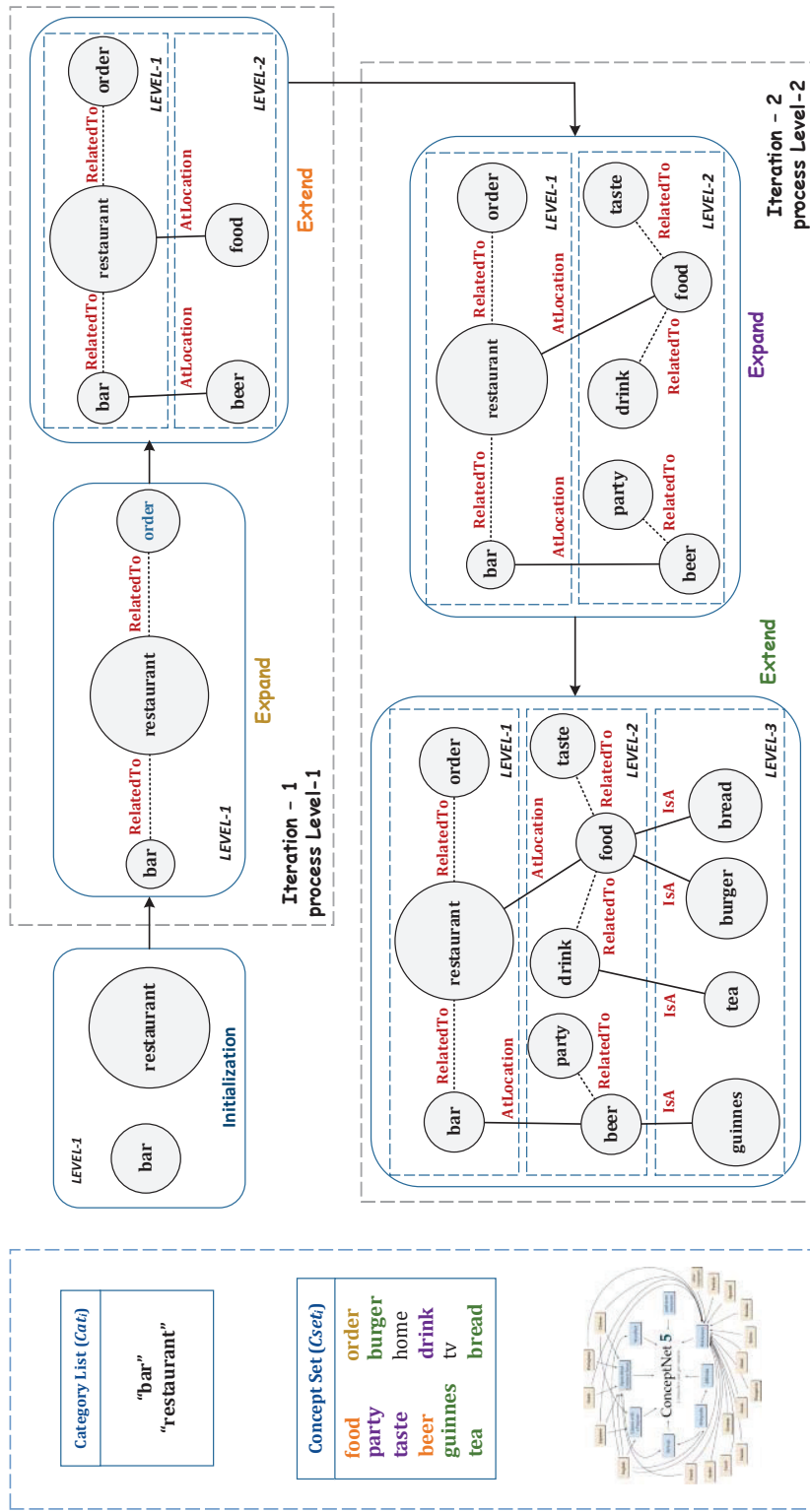

(a) Sample Inputs

(b) Evolution of CCH in Successive Iterations

Figure 4.5: CCH Formation Process.

Once the construction of the CCH is over, we discover relevant activities from $Aset_i$. We process each activity in $Aset_i$ and declare it as relevant if the concept associated with that activity is present in the CCH. In summary, given a set of activities, we discover the concepts associated with those activities, organize them into the CCH and use this hierarchy to infer relevant activities.

## 4.3 Sense-aware Non-Redundant Activity Discovery

The activities discovered in the previous stage are relevant but many of them may be redundant with respect the sense of an activity. For example, both the activities (*have, food*) and (*take, food*) are associated with the same concept "*food*" and hidden sense "*eat*" and so, they are redundant to each other. To handle the activity redundancy issue, we utilize the notion of hidden senses and discover the redundant activities (which are associated with a common hidden sense and have same concept) and merge them into a single activity. This process can be thought as the *sense-based activity clustering*, where each cluster is composed of a set of redundant activities. For example, considering the common hidden sense "*eat*", activities (*take, food*), (*get, food*) and (*have, food*) form a single activity cluster and is represented by (*have/get/take, food*) [see Figure 4.6(a)]. Similarly, example of another activity cluster is (*make/prepare, food*) formed based on the common hidden sense "*cook*" [see Figure see Figure 4.6(b)]. However, activities (*have, chicken*) and (*have, food*) have common hidden sense "*eat*", but as their concepts are different, they are not redundant and not merged.

Given a set of relevant activities $Aset_i$ for location $L_i$, we iteratively merge a pair of activities having common hidden sense and same concept. Note that, the concept of an activity is known, but the activity sense is unknown. Here, we use "*RelatedTo*", "*IsA*" and "*UsedFor*" relations of ConceptNet to discover the hidden senses associated with an activity. For example, the common hidden sense associated with activity (*take, food*) and (*have, food*) is "*eat*" based on the relationship $\{eat \overset{\text{RelatedTo}}{\rightarrow} take\ food\}$ and $\{have\ food \overset{\text{UsedFor}}{\rightarrow} eat\}$ respectively. So, both activities are merged into single activity (*have/take, food*).

Above procedure merges activities that are mostly associated with generalized concepts. For example, we have merged many activities that are associated with the concept "*food*". However, the procedure fails in most of the cases when activities are associated with specialized concepts. For example, redundant activities (*have, burger*) and (*take, burger*) are not merged using the above mentioned procedure. But, if we take the general-

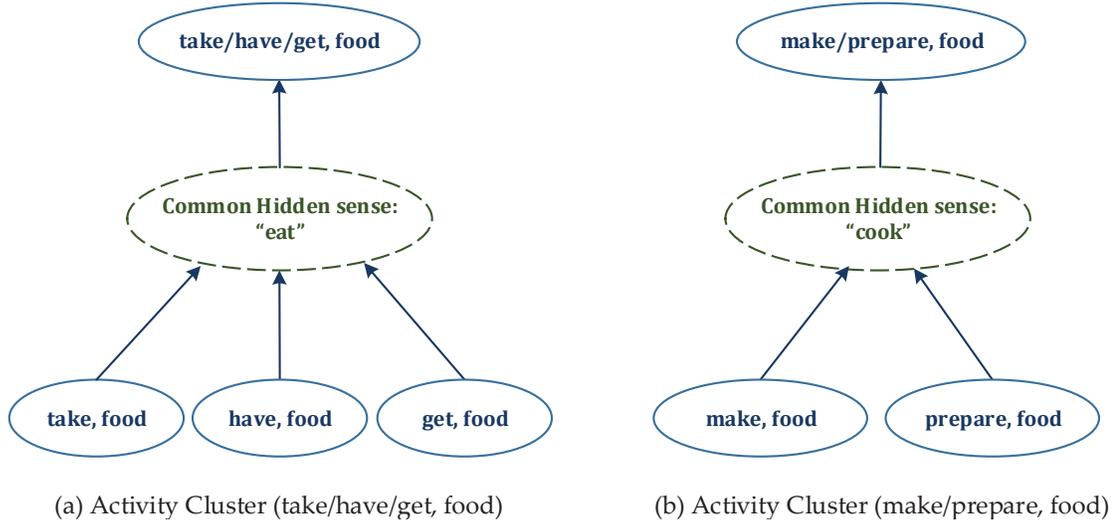

(a) Activity Cluster (take/have/get, food)    (b) Activity Cluster (make/prepare, food)

Figure 4.6: Examples of sense-based activity clustering.

ized concept of "*burger*" into account, we can merge (*have, burger*) and (*take, burger*) into a single activity. According to the CCH shown in Figure 4.5(b), the generalized concept of "*burger*" is "*food*". Since, (*have, food*) and (*take, food*) are already merged based on the common hidden sense "*eat*", we can also merge (*have, burger*) and (*take, burger*) based on the same hidden sense. This example suggests that we can apply the idea of utilizing generalized concept to merge the specialized activities. In summary, we iteratively merge two redundant activities having common concept $c$ if the activities associated with the *generalized concept of $c$* are already merged. In this iterative process, we reuse CCH to infer the generalized concept.

At the end of this phase, the discovered relevant and non-redundant activities in $Aset_i$ are ranked based on their $AF$ values and stored in a repository. Note that, merging of redundant activities increases their overall support (i.e., $AF$ value). This intern helps in obtaining correct activity ranking based on their $AF$ values and ensures generation of correct result(s) during processing of location-based skyline queries [26].

## LANet Formation

USEFULLNESS of Location-aware Activity Recommendation System depends on the enrichment of its back-end knowledgebase. In chapter 4, we discussed about location-specific activity discovery processs using a novel technique called `ActMiner`. But, discovering activities is not enough for activity recomeendation. We need an efficient and enriched knowledgebase where wide range of queries can run and desired results can be retrieved in real time.

In this chapter, we introduce the formation of `LANet`- the proposed unified and enriched knowledgebase for activity recommendation. The chapter is organized into six major section, each of which contribute to the sequential development of the proposed knowledgebase. In this first section, we discuss about how Location Nodes are formed with a set of well defined properties. Next, we discuss about the formation of Activity Nodes and Activity-location Links in `LANet`, followed by formation of Inter-Location Links that connects locations which supports similar activities. Then, we augment `LANet` with the concept of Boundary of Uniqueness of an activity which concerns about the uniqueness of an activity at a particular location. Finally, we present the overall representation of LANet in the form of a property graph model and how it is stored for query processing purpose.

## 5.1 Initialization Phase: Formation of Location Nodes with Location Node properties

In the initialization phase of the `LANet` discovery process, *Location Nodes* are created along with a set of properties. For each review set for a given location, we form a Location Node in `LANet`. Thus, if input to our `LANet` discovery process is $m$ number of review sets for $m$ locations, $m$ number of Location Nodes are created in `LANet`. During formation of Location Nodes, we augment each location Node with the following set of properties containing information about a location:

1. **Name_of_Location.** The name of the location for which the location node is created.

2. **Formatted_Address.** The address of the location formatted in the following pattern: *street_information, city, state, PIN, country*.

3. **Latitude.** The latitude value of the location.

4. **Longitude.** The longitude value of the location.

5. **Category.** List of categories representing the semantics of the location.

6. **No_of_Reviews.** No_of_Reviews tell about how many review messages have been written about that location which basically shows the popularity of that location among visitors. A location with higher number of review messages implies that lot of people have visited and liked that place.

## 5.2 Formation of Activity Nodes and Activity-Location Links

Once all the Location Nodes are created along with their property values, we proceed to discover the activities supported by those locations. To achieve the objective, we use `ActMiner` (see chapter 4) as an activitiy discovery module of `LANet` Formation process. Given the categories and review sets for all $m$ locations, `ActMiner` generates a set of *relevant* and *non-redundant* activities, denoted as $Aset_i$ for each location $L_i$ by following a sequential three phase Discover-Filter-Merge based technique and stores them in a *Gobal Activity Repository*, denoted as $Aset$. $Aset$ contains information about all the location-specific activities for all $m$ locations along with their $AF$ values.

Now, we extract a set of activities form *Aset* having distinct *Activity_Name* represented by (*verb, noun/noun phrase*) pair. For each distinct *Activity_Name* in the mentioned set, we create a *Activity Node* in `LANet` and link the node with all the Location Node where this activity is performed by a *Activity-Location Link*. We use *Aset* to infer the information about the locations where a particular activity is performed to aid the linking process. Each *Activity Node* is associated with only one property, i.e., **Activity_Name** as mentioned before. The *Activity-Location Link* between an activity Node and a Location Node in **LANet** is labeled with a relationship name "*Is_Performed_At*" (which tells about the fact that the activity is performed at that location) and is accompanied with a set of properties containing location-specific information about the activity at that location. Description these properties, their significance and calculation of their property values are discussed below.

1. **Activity_Frequency.** The AF value of an activity at a particular location (Recall Definition 3.2) and computed at the end of Dependency-aware Activity Extraction Phase of `ActMiner` and retrieved from *Aset* during link formation.

2. **Activity_Popularity_Index.** The Activity_Popularity_Index (API) value of an activity at a particular location tells about the relative popularity of an activity at that location compared to other locations where the activity is performed. The API value of an activity at a particular location is computed at the end of *Activity-based Location Similarity Detection Phase* (see section 5.3.1) and augmented with the corresponding *Activity-Location Link*.

3. **Generalized_Concept_Score.** The Generalized_Concept_Score ($GC_{score}$) value of the concept associated with an activity performed at a given location are used for generalized activity ranking.

4. **Specialized_Concept_Score.** The Specialized_Concept_Score ($SC_{score}$) value of the concept associated with an activity performed at a given location are used for specialized activity ranking.

## Concept Scores Calculation

The $GC_{score}$ and $SC_{score}$ values of an activity at a given location are used for activity ranking to facilitate specific query processing purpose. The $GC_{score}$ and $SC_{score}$ values are basically two types of *Concept Score* values of the Concept associated with an activity and act as a measure of the degree of specialty and generality of the concept as well as its associated activity respectively at a particular location. Depending on a "*weight*" value while calculating the *Concept Score*, the corresponding Concept score value is termed as Generalized Concept Score ($GC_{score}^{k}$) or Specialized Concept Score ($SC_{score}^{k}$).

In order to calculate these concept score values, we reuse CCH (see section 4.2) to infer the "*weight*" to quantify the degree of generality and specialty of an activity associated concept for a given location. Let, $m$ is the number of relevant concepts present in the CCH for location $L_i$. Now, given an activity $A_i^j \in Aset$, where $A_i^j$ is in the form of ($verb_j, concept_k$) pair, the concept scores of its associated concept "$concept_k$", is computed as shown in the equations 5.1 and 5.2 below.

$$(5.1) \qquad GC_{score}^{k} = \log CF_k \times \frac{1}{CLI_k}$$

$$(5.2) \qquad SC_{score}^{k} = \log CF_k \times CLI_k$$

where, $CF_k$ is the *Concept Frequency* of "$concept_k$" and $CLI_k$ is the *Concept Level Index* of concept "$concept_k$" and denotes the index of the level in which the relevant concept lies in the CCH, assuming the index of the level of root concepts as 1, i.e., if "$concept_k$" lies in level $l_r$ in CCH, $CLI_k$=r. As discussed in section 3, the more we move down the levels of a given CCH, the more we encounter specialized concepts rather than generalized concepts. Hence, the $CLI$ value of a given concept shows how much extent the concept is generalized or specialized and so is used as a weight for concept score calculation purpose.

## Role of Concept Scores in Query processing

Activities at a given location are ranked based on their *AF* values and concept scores acts a filtering parameter in the process of ranking which gives user the flexibility to issue specialized queries to `LANet`. If the user wants to get *top-k* activities that are associated with *top-m* specialized concepts, the query processing module first searches for *top-m* distinct concepts based on their $SC_{score}$ values. Next, the set of activities that are associated with those concepts are found out and then, ranking is done in the decreasing order of their *AF* values. Similarly, processing can also be done in the case of query like "*recommend top-k activities that are associated with top-m generalized concepts*" by considering $GC_{score}$ values for generalized activity ranking purpose.

5. **Boundary_of_Uniqueness.** The Boundary_of_Uniqueness (BoU) value of an activity at a given location gives the uniqueness information about that activity at that location compared to other locations where the activity is performed. Details about the BoU property and its procedure of computation is discussed in section 5.4.1. The BoU property values are augmented with the corresponding *Activity-Location Link* at the end of `LANet` discovery Process.

## 5.3 Inter-Location Link Formation

So far, we have discussed about how *activity nodes* are formed in `LANet` and gets associated with the corresponding locations by *Activity-Location Links* enriched with a set of properties. Now, in this section we introduce the concept of *Inter-Location Links* which connects two Location Nodes in `LANet` based on the concept of *activity-based location similarity*. These links are also associated with a set of properties that facilitates several spatial applications like alternate location finding, common activity detection etc. as discussed in chapter 1.

### 5.3.1 Activity-based Location Similarity Detection

In order to detect similarity between locations based on the location-specific activities performed there, we use the concept of *Term Frequency-Inverse Document Frequency* (TF-IDF) [20] and cosine similarity measure. TF-IDF is a very popular technique widely used for document similarity Detection. The basic idea of this numerical statistic is to find out the words or terms that are important to distinguish a document among a set of documents. As a document is described by terms, by estimating the frequency of their occurrences in the document, we can easily detect which terms are important and are responsible for distinguishing it from others.

The idea of applying TF-IDF and cosine similarity for the purpose of activity-based location similarity detection can be explained by using the analogy as shown in the Figure 5.1. According to the analogy, there exists a triangular relationship between location, location-reviews and location-specific activities where the location-reviews associated with a location act as a representative of that location and tell about location-specific activities which are performed at that location. So, here the location-reviews contain information about location-specific activities and how many times they are referred in the review document in the similar way any general document contain information about its terms and their occurrences where TF-IDF and cosine similarity is used for similarity detection.

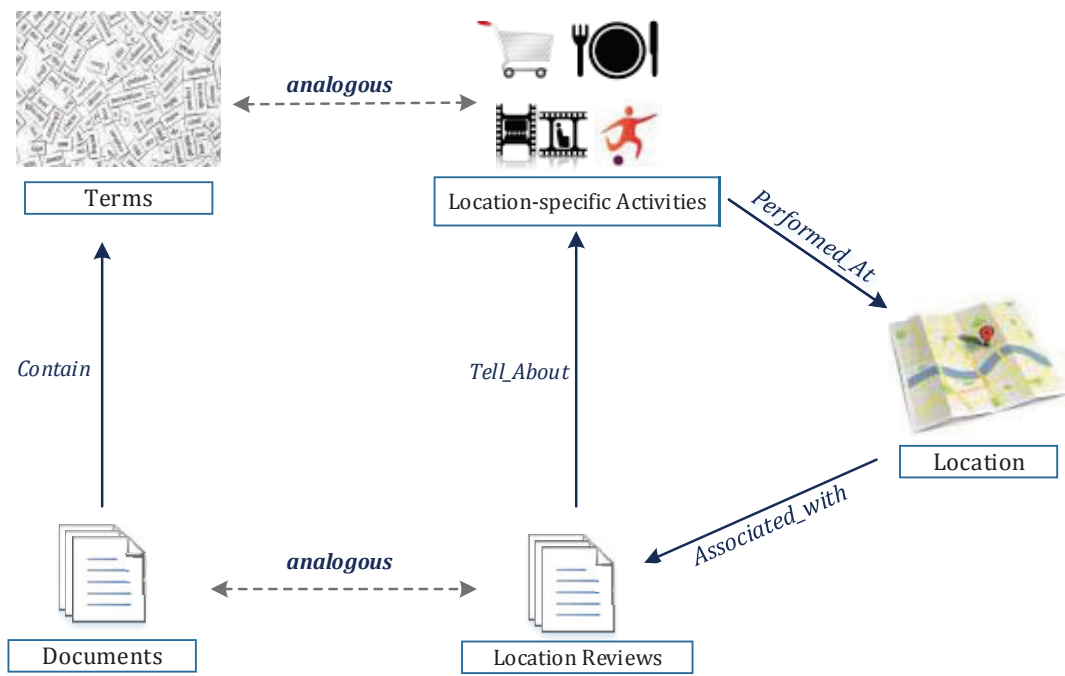

Figure 5.1: Analogy of locations, location reviews and location-specific activities with terms and documents.

As suggested in TF-IDF based document similarity approach, we prepare an *Activity-Location Matrix*(ALM) (see Figure 5.2), where each row represents one activity having unique *Activity_Name* which is denoted as $A_i$, where $1 \leq i \leq n$ , provided there exists $n$ activities with unique *Activity_Name*(s) in *ASet* and each column corresponds to a location from $L$. The element of the matrix $ALM_{ij}$ indicates the *Activity Frequency* of an activity $A_i$ performed at location $L_j$, the value of which is retrieved from *Aset* by searching the $AF$ value of the activity having same *Activity_Name* of $A_i$ in *Aset_j*. The Activity-Location Matrix is the key data structure to compute *Activity Frequency-Inverse Location Frequency* (AF-ILF) statistic just like TF-IDF in document similarity detection.

Once, the *Activity-Location Matrix* has been formed, the next step is to compute *AF-ILF vector* for each Location present in the matrix. To do this task, the AF-ILF values for each activity for each Location is computed where the AF value for each activity $A_i$ for each location is computed by following *logarithmically scaled Activity frequency* which is given by-

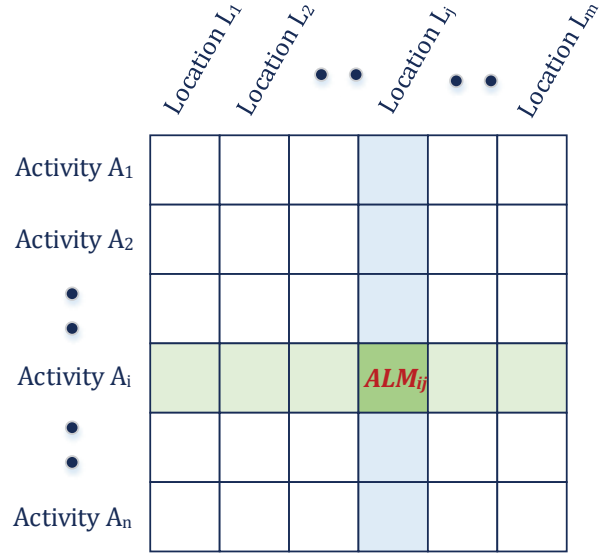

Figure 5.2: Activity-Location Matrix.

$$AF_{ij} = \log(1 + ALM_{ij})$$

And the ILF value for an activity $A_i$ is computed as:

$$ILF_i = \log_{10} \frac{|L|}{|\{ \, j \mid ALM_{ij} > 0 \, \}|} \qquad , where \, 1 \le j \le m$$

Next, the AF-ILF value of Activity $A_i$ at Location $L_j$ is computed as:

$$AFILF_{ij} = AF_{ij} \times ILF_i$$

In this way, our proposed solution computes the AF-ILF values for all activities. Now, let $\overrightarrow{L_k}$ be the AF-ILF vector formed using AF-ILF values for all activities for a given location $L_k$ present in the Activity-Location Matrix. So, what we get in this step is a set of AF-ILF vectors where each vector represents one location. These AF-ILF vectors for all locations are then used by *cosine similarity measure* to measure similarity between locations which is given by a statistic called *Similarity Index* (SI) value computed as shown below:

The *Similarity Index* (SI) value between Locations $L_p$ and $L_q$, denoted by $SI_{pq}$ is computed as given by the equation 5.3 shown below.

$$(5.3) \qquad SI_{pq} = \cos(\overrightarrow{L_p}, \overrightarrow{L_q}) = \frac{\overrightarrow{L_p} \cdot \overrightarrow{L_q}}{|\overrightarrow{L_p}||\overrightarrow{L_q}|}$$

We can explain the location similarity detection using *cosine similarity measure* on AF-ILF vectors using Figure 5.3 shown below. The idea behind this similarity detection is to map the AF-ILF vectors into a *n-dimensional* space where each dimension represents an activity present in ALM and the AF-ILF vectors are the representative of their corresponding locations lying in that n-dimensional space. Now, if we measure the Cosine of the angle between any two vectors, it gives the desired SI value between those two locations. If the angle between those two vectors are 90°, i.e., two vectors are orthogonal to each other, there is no similarity between those two locations and the SI value is 0. If the angle between them is 0°, then those two locations are exactly similar given by the SI value as 1. Otherwise, if the angle Θ is in between 0° and 90°, the two location has some positive similarity given by the corresponding SI value that lies between 0 < SI < 1. If the SI value between two locations is greater than 0, it implies that some common activities are performed at both locations and if it is 0, the two location have no activities in common.

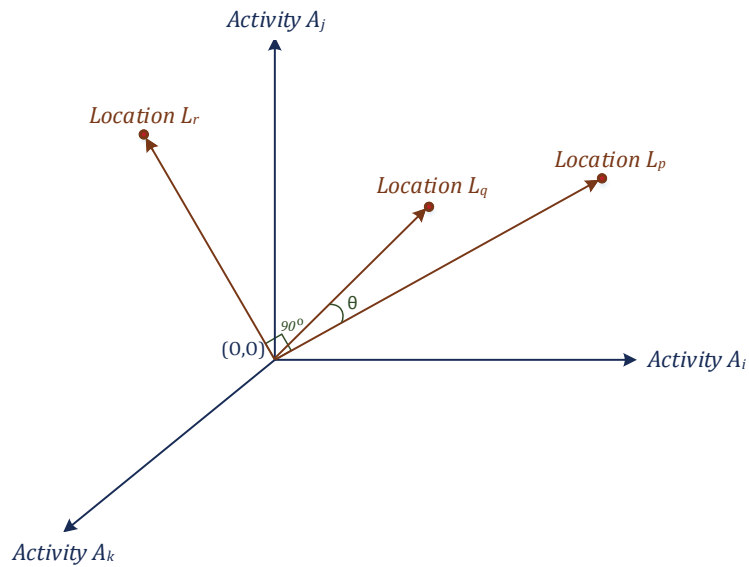

Figure 5.3: Applying Cosine similarity on AF-ILF vectors.

So, at the end of this step, we get SI values for all pair of locations. The results of this step is useful for calculating Boundary of Uniqueness (BoU)(Section 5.4.1) and building an application for finding an alternate location finding.

## Computation of Activity_Popularity_Index

As we have mentioned earlier, *Activity_Popularity_Index* specifies the relative popu-larity of an activity at a particular location compared to other locations where the activity is performed. The ALM matrix formed during *Activity-based Location Similarity Detection phase* is reused here to compute this relative popularity value. The *Activity_Popularity_Index* value of an activity $A_i$ at Location $L_j$, denoted as $API_{ij}$ is computed as given below.

$$API_{ij} = \frac{ALM_{ij}}{\sum_{j=1}^{m} ALM_{ij}} \qquad , where\, 1 \leq j \leq m$$

After computation of the API values, we augment the corresponding *Activity-Location Link* in `LANet` with the computed values. Thus, all *Activity-Location Links* in `LANet` get enriched with relative activity popularity information.

## 5.3.2 Augmenting LANet with Inter-Location Links

Once the processing of *Activity-based Location Similarity Detection* phase is over, we get the SI values for a set of pair of similar locations. For each pair of similar locations belonging to the mentioned set, we connect the Location Nodes corresponding to those two locations with a *Inter-Location Link*. Each *Inter-Location Link* is labeled with a relationship name "*Is_Similar_to*" (which tells about the fact that the connecting locations are similar to each other) and is accompanied with a set of properties (as described below) containing location-similarity information about the connecting locations.

1. **Similarity_Index.** The SI value of pair of locations connected by the *Inter-Location Link*.

2. **Common_Activity_List.** The Common_Activity_List (CAL) contains the list of common activities performed in both the locations connected by the *Inter-Location Link*. The CAL value is computed by taking intersection between the sets of activities supported by each of the linked pair of locations.

3. **Distance.** The radial distance between the linked pair of locations computed using their corresponding latitude and longitude values.

Thus, at the end of this phase, all the similar locations get connected in `LANet` with *Inter-Location Links* along with their associated property values which enables the processing of spatial queries likes alternate location finding and common activity searching etc issued by the user.

## 5.4 Empowering LANet with the Concept of Boundary of Uniqueness

At this stage, we have developed *LANet* to support all activity and location related queries. But, the Uniqueness information of an activity at a given location has not yet been introduced. In this section, we introduce the concept of uniqueness of an activity at a particular location which helps in inferring the spatial importance of that location. The following subsections presents the notion of Boundary of Uniqueness to quantify the uniqueness property of an activity at a particular location and describes how we have incorporated the discovered knowledge into the `LANet`.

### 5.4.1 Boundary of Uniqueness (BoU) Calculation

In this sub section, we discuss the procedure to calculate *Boundary of Uniqueness*(BoU) value for each of the recognized activities to find out the region within which a given activity is considered as unique. The definition of *Boundary of Uniqueness* is given below.

**Definition 5.1. Boundary of Uniqueness.** The *Boundary of Uniqueness* of an activity $A_i^j$ performed at location $L_i$ , denoted by $\text{BoU}(A_i^j)$ is defined as the radial distance around location $L_i$ that covers a circular area within which the activity $A_i^j$ is performed at no other location expect location $L_i$.

For example, Consider the Figure 5.4 shown below where we have shown 3 similar locations in Roorkee. These 3 locations, viz., "*Hotel Royal Palace*", "*Hotel Center Point*" and "*Sagar Hotel & Restaurant*" are very popular restaurants here. In the diagram, we have denoted these places with red, blue and green colours and also shown some eating activities that are frequently done there by common people. If we carefully observe the Figure 5.4, we can see that "*Sagar Hotel & Restaurant*" is at a radial distance of 100m from "*Hotel Royal Palace*" and the common activities performed there are (*eat, chicken*), (*have, butter nun*) and (*take dessert*). Similarly, "*Hotel Center Point*" is at a radial distance of 40m form "*Hotel Royal Palace*" and the common activities between them are (*have, butter nun*) and (*take, dessert*). So, if we consider the activities performed at "*Hotel Royal Palace*", the BoU value of (*eat, chicken*) is 100m, because the nearest location where this activity is done, i.e., "*Sagar Hotel & Restaurant*" is at a radial distance of 100m from "*Hotel Royal Palace*" and the activity is unique within the circular area (shown in figure by light green shade) with radius 100m centering "*Hotel Royal Palace*". Similarly, the BoU value of (*have, butter nun*) and (*take, dessert*) are 40m, because the nearest location where this activity is done, i.e., "*Hotel Center Point*" is at a radial distance of 40m from "*Hotel Royal Palace*" and nowhere these activities are done within the circular area shown by light blue shade in the Figure 5.4. The activity (*get, beer*) is a unique activity that is only done at "Hotel Royal Palace" and nowhere else considering these three locations.

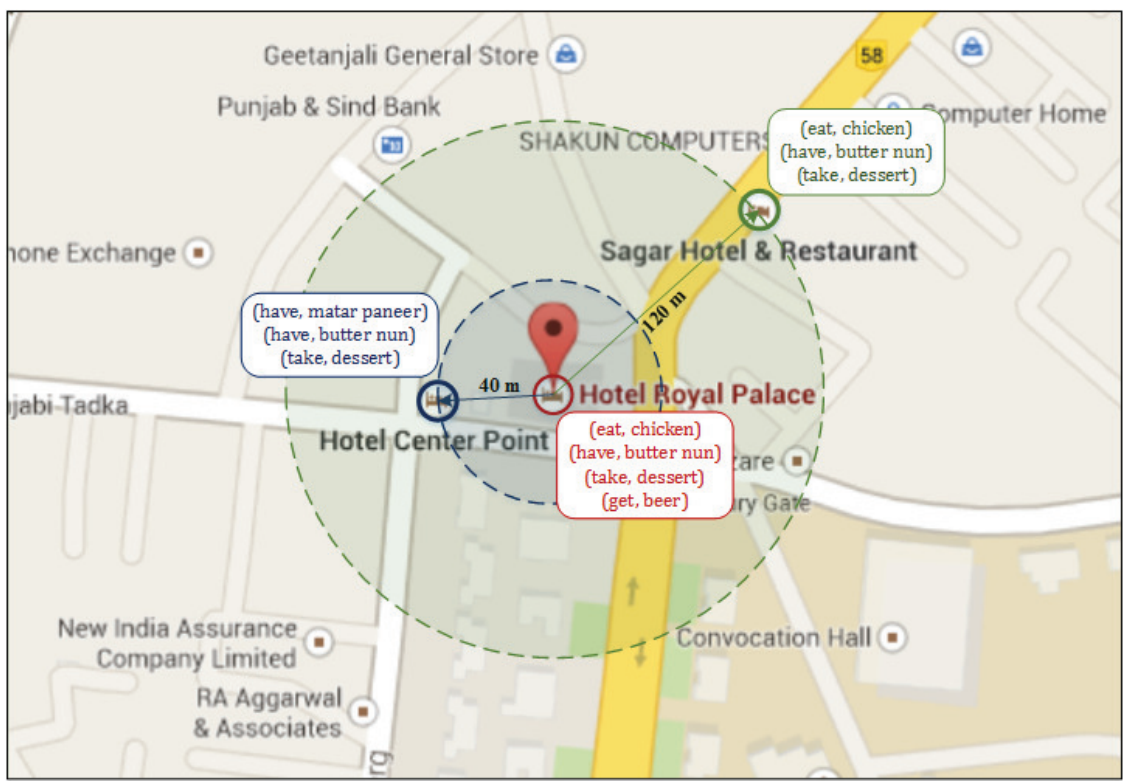

Figure 5.4: Diagram showing Boundary of Uniqueness and similarity between locations along with distance.

In order to calculate the BoU value of an activity at a particular location, we need to identify the nearest locations where that activity is performed. In absence of indexing techniques, the task of identifying the locations with same activity is computationally intensive as we have to scan all other locations. Since, we have already identified the SI values between any two locations, we can find out those locations having SI value > 0 with respect to the given location. If SI value between any two locations is greater than 0, it implies that both locations have at least one activity in common. In this way, we reduce the search space by eliminating the locations where the given activity is not performed at all. In addition to that, we have explored these locations having SI value>0 in the increasing order of their distances from the given location. This further helps us not to explore the locations that are far.

In summary, for calculating BoU value of activities at a particular location $L_i$, current phase first finds out the similarity set of $L_i$ that contains locations having positive SI value with $L_i$. Then, the phase calculates the radial distance between $L_i$ and each location from similarity set by using their latitude and longitude values and sorts them in increasing order of their distances. Next, the step selects the nearest location of $L_i$ from similarity set, finds out common activities between them and assigns the BoU value of each of those activities performed at $L_i$ to the radial distance between $L_i$ and the nearest location. Then, the step iteratively selects the 2nd, 3rd, ...., nth nearest location of $L_i$ and performs the same operations for remaining unassigned activities of $L_i$. This process terminates when all the locations in the similarity set have been accessed or all activities supported by $L_i$ have been associated with their BoU values.

In this way, the above process traces all the locations one by one and assigns BoU values of each of the activities associated those locations.

### 5.4.2 Augmentation of Activity-Location Link with BoU Property values

After discovering the knowledge of uniqueness of activities in terms of their BoU values, we incorporate these discovered knowledge into `LANet`. Each *Activity-Location Link* in `LANet` is augmented with the calculated BoU value of the activity for a given location. Thus, the `LANet` formation process ends and finally, we get a unified enriched knowledge-base for an efficient Location-aware Activity Recommendation System (LActRS) which is capable of supporting wide range of activity-related spatial queries form user.

## 5.5 Representation of LANet

In this section, we present the schema of the `LANet` which shows how the discovered knowledge is integrated in `LANet` in the form of a property graph [25] data model. Such a graph is useful in representing `LANet` and deploying the knowledgebse into a graph database format. This intern enables efficient query processing facility required by different location-aware recommendation applications. Figure 5.5 shows the final generic representation of `LANet` consisting of Location and Activity Nodes, Activity-Location Links and Inter-Location Links.

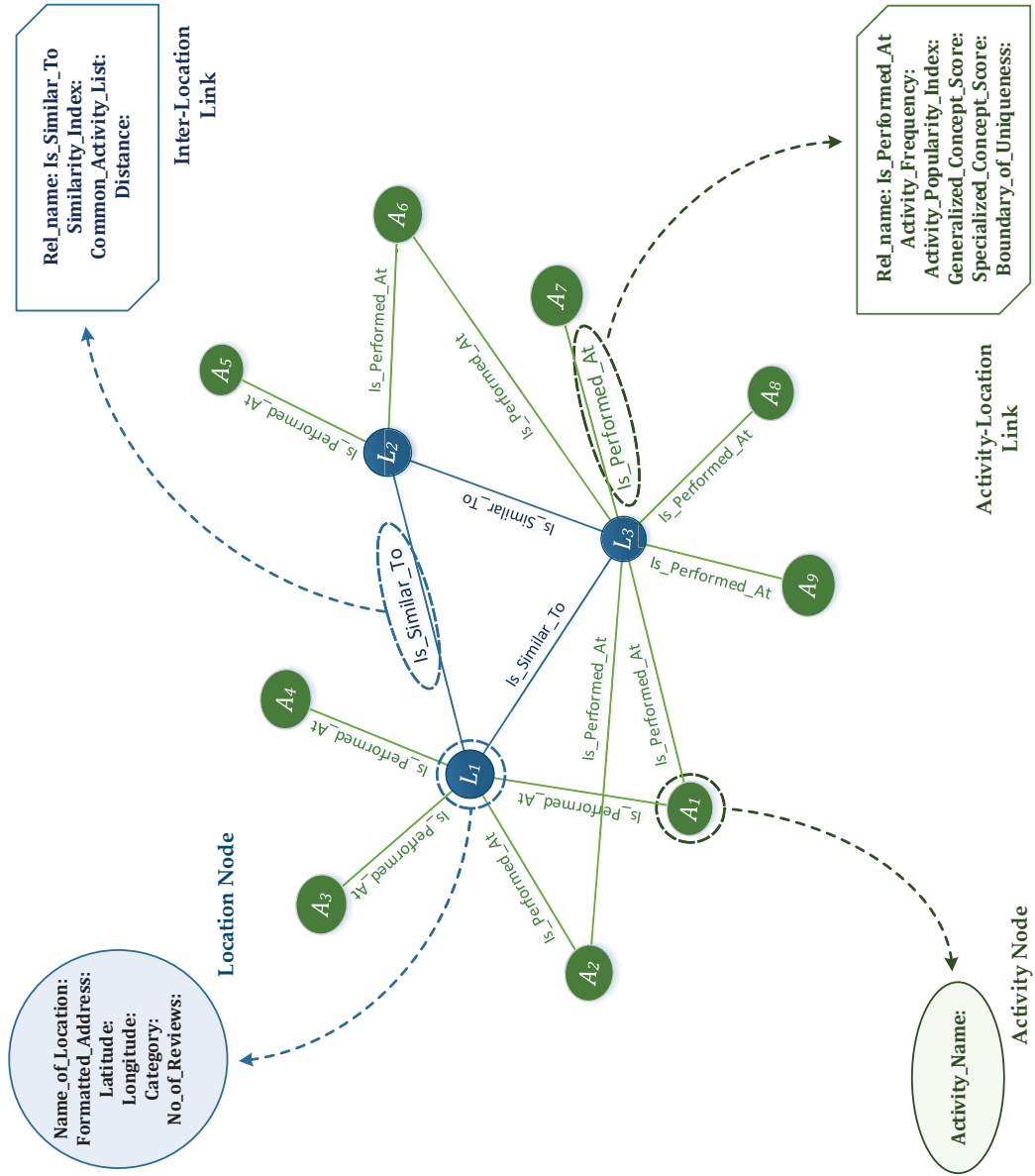

Figure 5.5: Diagram showing the generic structure of LANet as a property graph.

## 5.6 LANet As Graph Database

Graph Database [25] store data in the form of a graph and provides an elegant way to represent data and perform query processing on it. It is the most desirable solution to deal with large scale graphs where faster graph traversal is needed to retrieve information for application that requires real-time support. Examples of Graph databases that support "*property graph*" as their graph data model are Neo4j [3], Titan [6], OrientDB etc.

To employ `LANet` as a back-end knowledgebase of `LActRS`, we have used Neo4j for deploying `LANet` as a graph database. Contracting `LANet` as a graph database has allowed us to use graph database query languages Cypher (Neo4j) [3] to retrieve the required knowledge from `LANet` as per the query issued by the spatial applications in real-time.

EXPERIMENTAL EVALUATION

I N this chapter, we present the results of our experimental evaluation of `LANet` discovery process on two real world datasets. Our experiments evaluate `LANet` in term of its richness, accuracy of the knowledgebase, and practical usefulness in building real world applications. We use Java programming language to implement all the techniques presented in this paper. Table 6.1 shows the various libraries and knowledgebases we have used for implementing the functionality of the proposed techniques. We have also implemented the baseline approach [11] (see chapter 2) for comparison purpose.

## 6.1   Datasets

In our experimental evaluation, we have utilized publicly available yelp review dataset and collected one local Roorkee location review data set.

**Yelp Dataset.** This dataset contains 229,907 reviews from 43,873 users about 11,537 locations from the greater Phoenix, AZ metropolitan area. We choose 7 locations having more than 200 reviews and prepare a reviews set for each of the selected locations. Figure 6.1 provides the details of these locations along with their category information.

**Roorkee Dataset.** We have also collected a small set of review data for 25 locations in Roorkee. These locations are frequently visited by many students and institute staffs from IIT-Roorkee. In total, this local data set contains 686 review messages for 25 different locations as show in the Figure 6.2. Since, we are aware about our locality, this small dataset enables us to easily verify the results of our experimental studies.

Table 6.1: Various Libraries and Knowledgebases used in LANet implementation

| Purpose | Library/Knowledgebase |
|---|---|
| Sentence Extraction | Opennlp SentenceDetector (library) |
| Verbs, Nouns and Noun Phrase extraction | Stanford Part-Of-Speech Tagger (library) |
| Activity Extraction | Stanford Typed dependency parser (library) |
| Conversion of words into their base form | Java API for WordNet Searching (JAWS) (library) and WordNet (knowledgebase) |
| Relevant and non-redundant Activity Discovery purpose | ConceptNet (knowledgebase) |
| Knowledge Representation | Neo4j graph database |

| Loc_ID | Location Name & Address | Categories | No. of Reviews |
|---|---|---|---|
| 1 | **960 W University Dr** Tempe, AZ 85281, USA. | Pubs", "Bars", "Nightlife", "Restaurants" | 575 |
| 2 | **2611 N Central Ave**, Phoenix, AZ 85004, USA. | "Steakhouses", "Restaurants", | 278 |
| 3 | **401 E Jefferson St,** Phoenix, AZ 85004, USA. | "Arts & Entertainment", "Stadiums & Arenas" | 216 |
| 4 | **5701 N Echo Canyon Pkwy**, Phoenix, AZ 85073, USA. | "Active Life", "Climbing", "Hiking", "Parks" | 210 |
| 5 | **7107 E McDowell Rd**, Scottsdale, AZ 85257, USA. | "Food", "Sandwiches", "Breweries", "Pizza", "Restaurants" | 232 |
| 6 | **Galvin Bikeway**, Phoenix AZ 85008, USA. | "Arts & Entertainment", "Botanical Gardens", "Music Venues", "Nightlife" | 260 |
| 7 | **1514 N 7th Ave**, 2nd Fl, Phoenix, AZ 85007, USA. | "Bars", "Nightlife", "Lounges" | 232 |

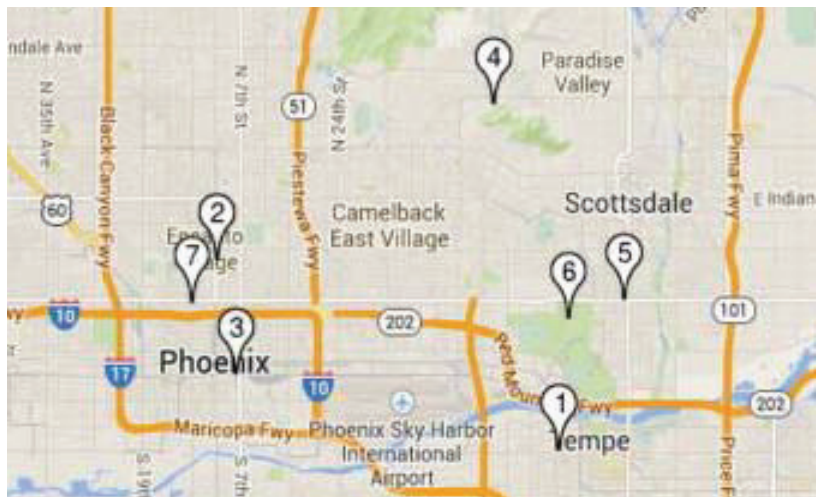

Figure 6.1: Selected Locations from Yelp Data set.

| Loc_ID | Location Name | Categories | No. of Reviews |
|--------|---------------|------------|----------------|
| 1 | Prakash Sweets. | "Dessert Shop","Snacks" | 61 |
| 2 | Kundan Sweets. | "Dessert Shop","Snacks" | 28 |
| 3 | Prakash Hotel. | "Hotel", "Restaurant" | 55 |
| 4 | Hotel Royal Palace. | "Hotel", "Restaurant","bar" | 38 |
| 5 | Dominos Roorkee. | "Pizzaries" | 46 |
| 6 | Sizzlers. | "Restaurant" | 13 |
| 7 | Food Point. | "Restaurant" | 19 |
| 8 | Motel Polaris. | "Hotel", "Restaurant" | 28 |
| 9 | NEEDS. | "Convenience Store" | 30 |
| 10* | The Pentagon Mall. | "Shopping Mall" | 35 |
| 11* | Vishal Mega Mart. | "Shopping Mall" | 22 |
| 12 | Woodland Exclusive Store. | "Garment Shop" | 18 |
| 13* | Reebok Store. | "Shoe Store" | 16 |
| 14 | The Raymond Shop. | "Suits", "Garments shop" | 14 |
| 15* | Nature Park. | "Park", "Hiking" | 09 |
| 16* | Solani Park. | "Park", "Hiking" | 09 |
| 17* | Crystal World. | "Water Park" | 16 |
| 18 | Hobbies Club. | "club","Recreation" | 26 |
| 19 | NESCAFE@IIT Roorkee. | "Cafe","Sancks" | 41 |
| 20 | Alpahar Canteen. | "Cafe", "Snacks" | 37 |
| 21 | Mahatma Ghandhi Central Library. | "Library" | 35 |
| 22 | Sports Complex. | "Sports" | 19 |
| 23 | PNB/SBI Bank. | "Bank" | 27 |
| 24 | Computer Centre. | "Cyber cafe","Computer" | 21 |
| 25* | Railway Reservation Centre. | "Ticket Reservation" | 23 |

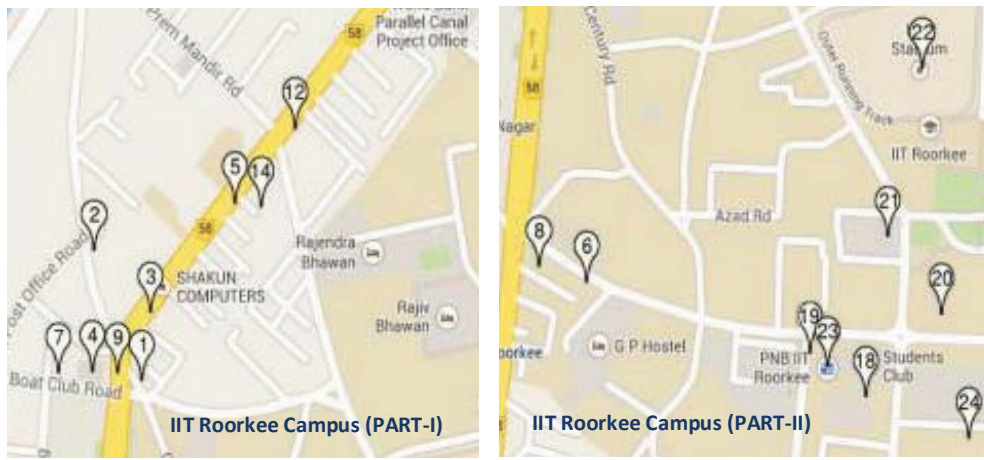

IIT Roorkee Campus (PART-I)  IIT Roorkee Campus (PART-II)

*\* These locations are not shown in the map due to space limitations*

Figure 6.2: Summary of Roorkee Dataset.

## 6.2   Experimental Study

In order to study the performance of our proposed solution to build `LANet`, we have evaluated the proposed solution based on the following evaluation parameters.

### 6.2.1   Evaluation 1: Information Richness and Correctness of LANet

In this subsection, we present the information richness of `LANet` and evaluate its correctness. First, we have discussed about some knowledgebase statistics about `LANet` formed using both the Yelp and Roorkee dataset. Then, we evaluate the correctness of the information stored in `LANet` based on three parameters, viz. *Location Similarity*, *Activity Popularity* and *Activity Uniqueness*. The details of the experiments done for the said evaluation purpose are presented below.

**Knowledgebase Statistcs.** We learn `LANet` from both data sets and analyze the richness of the developed knowledge base in inferring the information about location-specific activities. The `LANet` developed using yelp data set contains 7 location nodes, 6256 activity nodes and total 7270 links and took around 5.01 MB storage space. Similarly, the `LANet` formed using Roorkee dataset contains 25 location nodes, 886 activity nodes and 1250 links with a storage space of 2.59 MB.

Figure 6.3 presents a snapshot of a part of the `LANet` discovered having Roorkee locations of ID **1** to **7**. Each location node is marked with green color and activities associated with locations are marked with red colour. The *Activity-Location Links* are shown by red coloured lines with label "*Is_Performed_At*" connecting various activities with their corresponding locations. The *Inter-location links* are represented by green coloured lines with label "*Is_Similar_To*" showing similarity between locations. The node and link properties are not shown in the figure due to space constraints.

**Evaluation of Location Similarity.** In our first set of experiments, we prepare heat maps to captures the similarity between locations. Our hypothesis is that similar activities should be performed on locations with similar categories. Careful observation of Figure 6.4 shows that location ID-1 exhibits highest similarity to location ID-2 and second highest similarity with location ID-5. Now, from Figure 6.1, we see that the categories of these 3 locations are "*restaurants*". So, lot of activities are common in between them compared to the other locations.

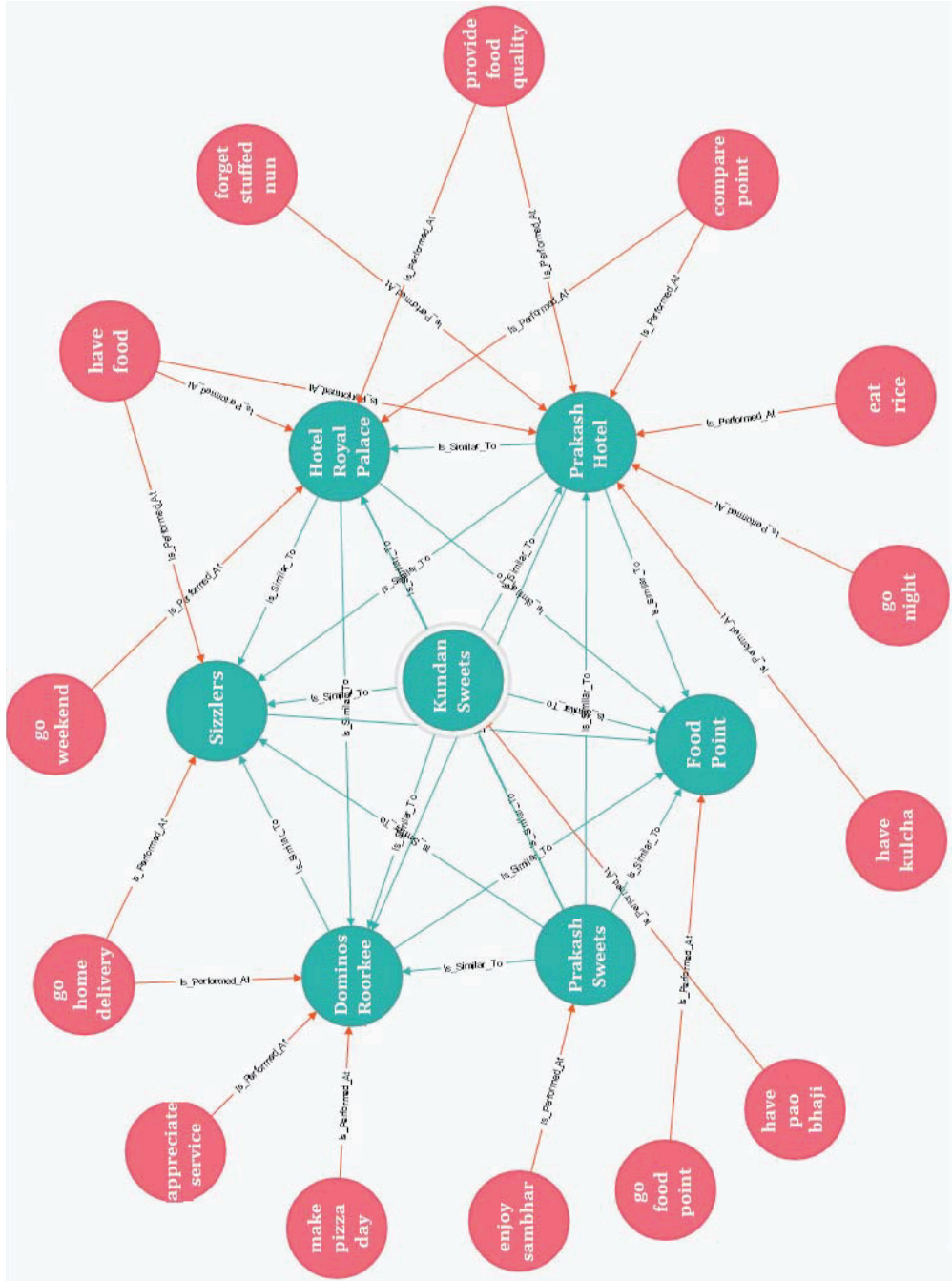

Figure 6.3: Snapshot of a part of the LANet discovered from Roorkee Dataset having location ID- 1 to 7 with various location-specific activities.

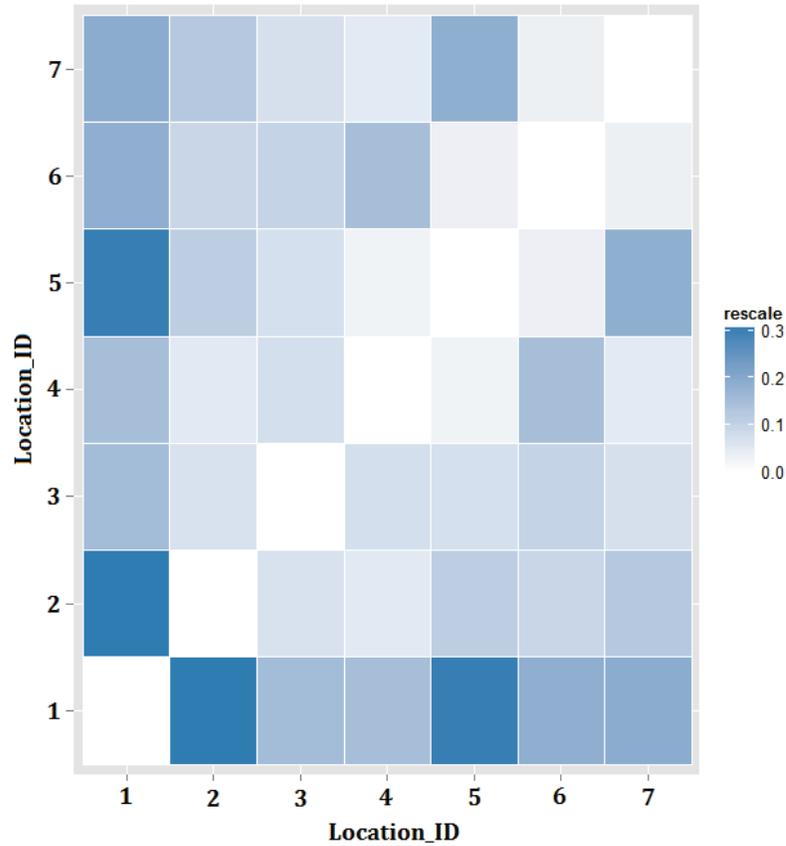

Figure 6.4: Heat map showing location similarity on yelp dataset.

Similarly, in the Figure 6.5, we have shown the similarity among 25 locations of Roorkee dataset. Here, location ID-19 and location ID-20 show the highest similarity in between them. These two locations are situated within our IIT campus and are well known for having tea, coffee and snacks etc. Again, location ID-3 and ID-4 of Roorkee dataset shows second highest similarity as shown in the Figure 6.5. Both these locations are well-known restaurants in Roorkee and frequently visited by local people for having dinner. Therefore, from this analysis, we can conclude that the results we have obtained, agree with the facts of the real-world.

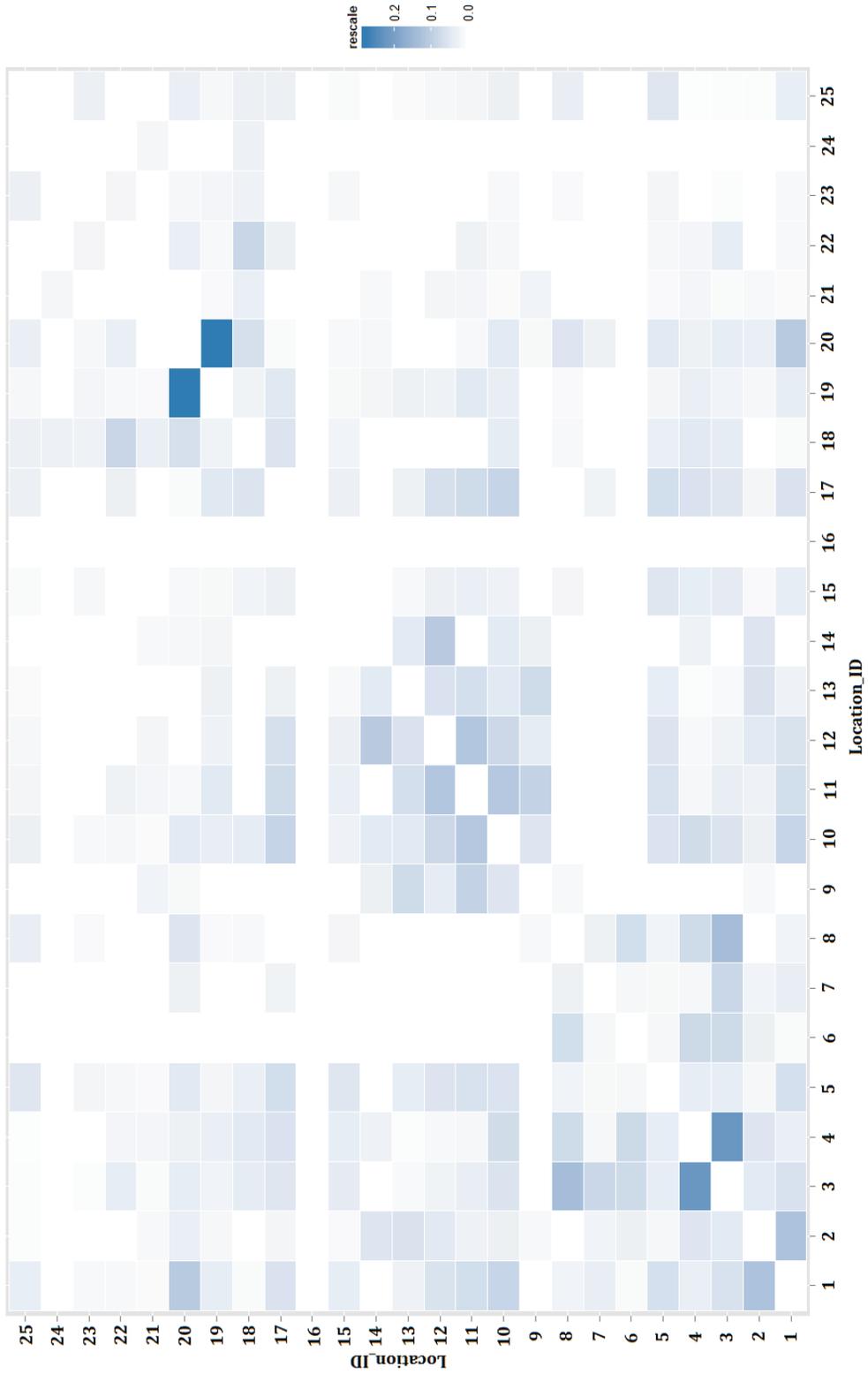

Figure 6.5: Heat map showing location similarity on Roorkee dataset.

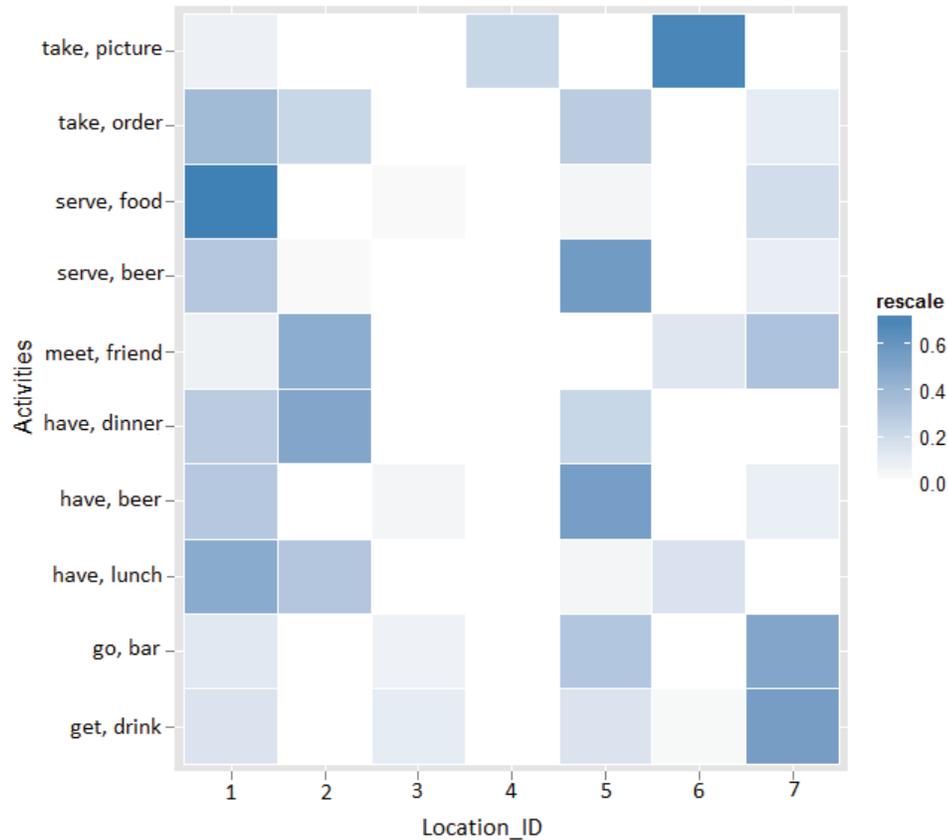

Figure 6.6: Popularity of Inferred activities on Yelp dataset.

**Evaluation of Activity Popularity.** In our second set of experiments, we prepare heat maps to compare the popularity of a set of activities among various locations. Figure 6.6 shows the heat map of activity popularity of 10 different activities that are performed at more than one location in yelp data set. Here, we can see that activity (*take, picture*) is most popular at location ID-6. And from the Figure 6.2 location ID-6 is a "*botanical garden*" by its category. Similarly, (*serve, food*), (*sit, bar*), (*have, experience*) and (*have, lunch*) activities are mostly popular in location ID-1 which is a "*restaurant*" cum "*bar*".

From the analysis of the results from Roorkee data set shown in the Figure 6.7, the activity (*eat, pizza*) has got the highest popularity at location ID-5 which is a "*pizzeria*". Activity (*recommend, veg food*) has got the highest popularity in location ID-3 which is the most well known "*hotel*" cum "*restaurant*" that sells vegetarian foods in Roorkee. Apart from that, the heat map shows location ID-1 as the most preferable place for having Indian desserts and snacks like rashmalai, samosa etc. in Roorkee. All these results are also valid considering the real life scenario.

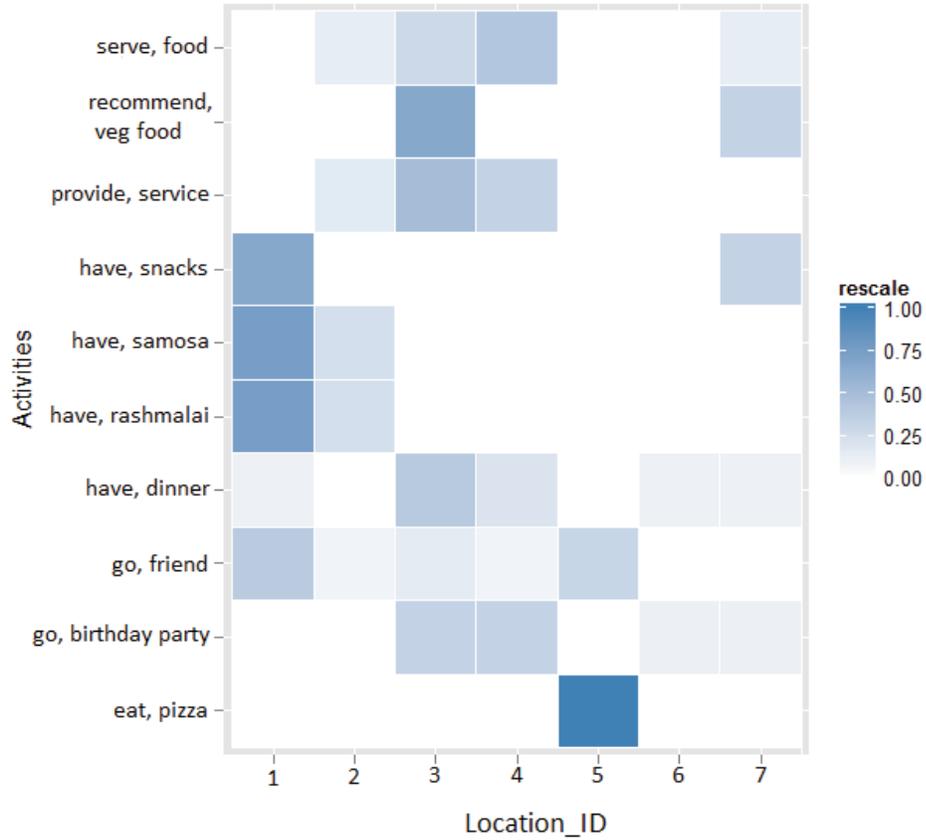

Figure 6.7: Popularity of Inferred activities on Roorkee dataset.

**Evaluation of Activity Uniqueness.** In our last set of experiments to evaluate the correctness of `LANet`, we have generated some statistics about activity uniqueness for analysis purpose. Table 6.2 presents the statistics of uniqueness of 7 activities performed at Location_ID-1 considering Yelp dataset. In this table, we have shown, the list of alternative locations where a given activity is performed apart from Location_ID-1, presented in the increasing order of their distance. Form this list of alternative locations, it is quite clear that the nearest location where an activity is performed is the first location in the list and the BoU value will be its distance from Location_ID-1.

In Table 6.2, we can observe that, Location_ID-5 is the nearest alternative location where (serve, beer) activity is performed which is situated at a distance of 5.24Km from Location_ID-1. The other locations having ID 2 and 7 are situated further form Location_ID-1 compared to 5 and so, with in a circular region of radius 5.24Km with center at location with ID-1, there is no other locations where one can get beer. We have evaluated the boundary of uniqueness of activity by considering only the 7 locations that belongs to the datset we have used for experimental analysis. In reality, there may be a location within the specified region where the activity is performed. But, with respect to these 7 locations, the activity is unique within a radial distance of 5.24 Km. Other activities and their uniqueness statistics also bears similar significance as shown in the Table 6.2.

Table 6.2: Statistics of uniqueness of 7 activities performed at Location_ID-1 (Loc: *960 W University Dr Tempe, AZ 85281, USA.*) considering Yelp Dataset.

| Activities | List of Alternative Locations Ordered with respect to Distance | BoU value | Nearest Alternative Location |
|---|---|---|---|
| (serve, beer) | 5, 2, 7 | 5.24Km | 5 |
| (have, experience) | 6, 3, 4, 7 | 4.24Km | 6 |
| (have, steak) | 2 | 12.84Km | 2 |
| (go, lunch) | 5, 2 | 5.24Km | 5 |
| (take, order) | 5, 2, 7 | 5.24Km | 5 |
| (have, party) | 6, 2, 7 | 4.24Km | 6 |
| (recommend, cheese) | 2 | 12.84Km | 2 |

In Section 6.2.2, we have evaluated the accuracy of the activities that take part in `LANet` formation process. As we can observe in the Section 6.2.2 that our activity discovery method `ActMiner` is quite accurate, we can claim that the BoU values computed by our approach are also accurate and proves the efficacy of our uniqueness computation method.

Apart from yelp dataset, we have also used Roorkee dataset for collecting statistics about the uniqueness information of the activities supported by Roorkee locations. Table 6.3 presents the statistics of uniqueness of 7 activities performed at Location_ID-4 considering Roorkee Dataset. Being a local dataset, we have verified these results and found that the results conform to the reality.

In summary, from the analysis of `LANet` based on various perspectives as discussed above, we can conclude that `LANet` has successfully captured the information about location and its supported activities, which also conforms to the real-life beliefs of local people and thus, it proves the correctness of our proposed solution.

Table 6.3: Statistics of uniqueness of 7 activities performed at Location_ID-4 (Loc: *Hotel Royal Palace, Civil Lines, Roorkee.*) considering Roorkee Dataset.

| Activities | List of Alternative Locations Ordered with respect to Distance | BoU value | Nearest Alternative Location |
|---|---|---|---|
| (have, dinner) | 7, 1, 3, 8, 6 | 51.45 m | 7 |
| (go, friend) | 1, 3, 2, 5, 12, 13, 11, 19, 25, 17, 10, 15 | 77.01 m | 1 |
| (serve, food) | 3, 2 | 118.157m | 3 |
| (go, birthday party) | 3, 8, 6 | 118.157m | 3 |
| (spend, time) | 1, 3, 5, 20, 15, 19, 18, 20, 10 | 77.01 m | 1 |
| (have, fun) | 3, 18, 17, 10 | 118.157m | 3 |
| (provide, food quality) | 3 | 118.157m | 3 |

## 6.2.2 Evaluation 2: Measuring Accuracy of ActMiner.

In order to measure the accuracy of discovered activities, we manually obtain the list of possible activities for each location. We have distributed review set among 54 master students in our institute and ask them to write down the activities in terms of (*verb, noun/noun phrase*) pairs along with the review message id. We consider these activities as the ground truth. Let, the set of activities inferred using human perception for a given Location $L_i$, be $GT_i$, i.e. ground truth for location $L_i$ and the set of activities discovered by ActMiner is $Aset_i$. Then, the accuracy of ActMiner for location $L_i$ is calculated as follows-

$$Accuracy_i = \frac{|Aset_i \cap GT_i|}{|Aset_i|}$$

where, $(Aset_i \cap GT_i)$ gives the set of activities discovered by ActMiner which also have been inferred by human perception and so, is considered as the set of correct and meaningful activities. The set of Activities $(Aset_i - GT_i)$ are the incorrect ones. This set contains activities that are either not meaningful in sense or not being detected using human perception while analyzing the corresponding reviews from which it has been discovered by ActMiner.

**Accuracy of ActMiner on Yelp dataset.** For comparative study, we obtain the set of activities using baseline and two versions of ActMiner. ActMiner-1 discovers activities using dependency-aware activity extraction technique, whereas ActMiner-2 discovers activities using the idea of dependency-aware and category-aware activity discovery techniques. We have not incorporated the third, i.e., "*Merge*" phase of ActMiner for accuracy evaluation purpose as redundancy minimization does not effect the accuracy. On Yelp data set, we discover top 500 activities for each location using ActMiner-1,

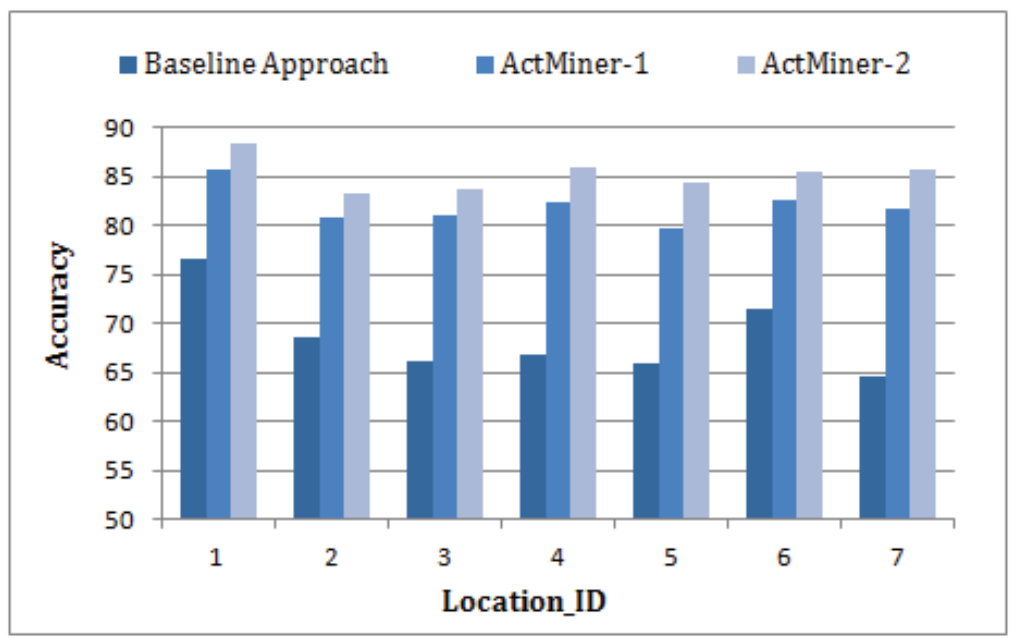

Figure 6.8: Accuracy of Inferred activities on Yelp dataset

`ActMiner-2` and Baseline. Figure 6.8 shows that `ActMiner-1` outperforms the baseline method in terms of accuracy. More specifically, using Yelp data set, on an average, the baseline approach has achieved accuracy of 68.6% considering all 7 locations whereas dependency-aware activity extraction method, i.e., `ActMiner-1` has obtained an average accuracy of 82% showing significant improvement of 13.4%. Moreover, considering top 500 relevant activities discovered using category-aware relevant activity discovery approach, `ActMiner-2` has obtained an average accuracy of 85.23% which implies 3.23% average improvement in accuracy to that obtained using `ActMiner-1`.

**Accuracy of ActMiner on Roorkee dataset.** On Roorkee data set, we discover all activities for each location using `ActMiner-1` and Baseline. Figure 6.9 shows the comparison of accuracies for both the approaches. Again, we observe that the activities discovered by `ActMiner-1` are more accurate. In terms of statistics, on an average, the baseline approach has achieved accuracy of 74.787% considering all 25 locations whereas `ActMiner-1` has obtained an average accuracy of 83.728% showing significant improvement of 8.94% in accuracy compared to the baseline approach. We have not obtained the accuracy of `ActMiner-2` on Roorkee dataset as the reviews in this dataset contain local or Indian concepts that are mostly not available in the ConceptNet and hence, are not suitable for the relevant activity detection purpose.

In summary, we can say that `ActMiner` performs more accurately than the baseline approach and hence ensures the accuracy of inferred activities which take part in building `LANet`. The reason behind this is that we have used the dependency relations between words as the metric for activity extraction which ensures the pairing of words to be meaningful. Apart from that, most of the irrelevant activities thrown out after category-aware relevant activity discovery phase, cause improvement in accuracy in `ActMiner-2`.

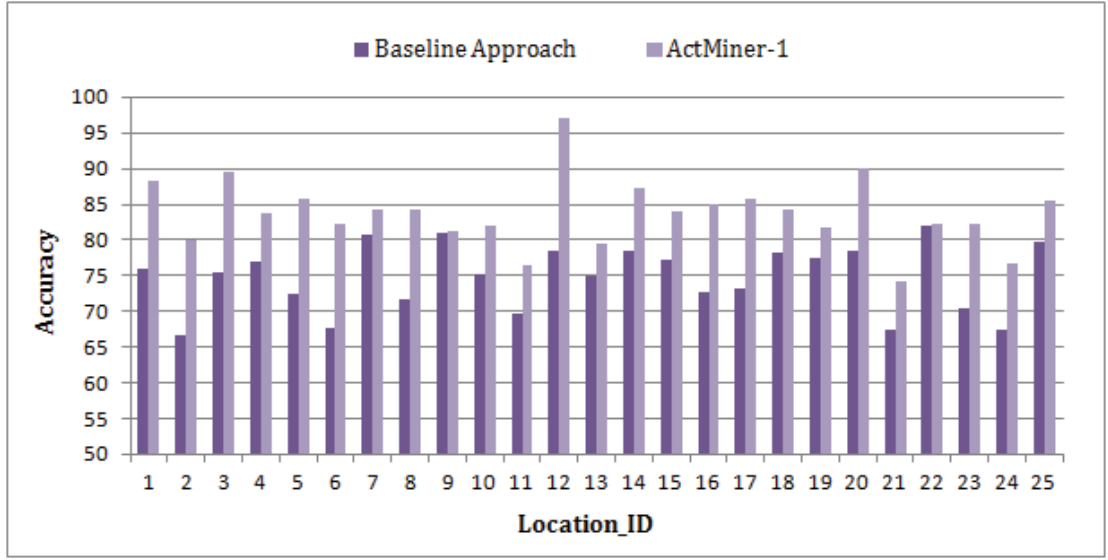

Figure 6.9: Accuracy of Inferred activities on Roorkee dataset.

So far, we have evaluated our system based on the richness and accuracy of the discovered activities. Now, we present the qualitative analysis of `LANet` from application point of view. In this Analysis, we have conducted experiments to verify the performance of our system in *Broadcast Service* which concerns about the redundancy elimination issue, activity ranking for skyline query processing and performance in activity recommendations as discussed below.

## 6.2.3 Evaluation 3: Qualitative Analysis I: Broadcast Service.

Now, we investigate the usefulness of sense aware redundant activity discovery phase in terms of advertising popular activities in broadcast environment. We manually obtain the list of redundant activities from the output of `ActMiner-2`. We consider these set of marked activities as the ground truth for redundancy checking purpose. Next, we run the process of sense-aware non-redundant activity discovery on the output of `ActMiner-2`. Figure 6.10 shows the number of redundant activities before and after the sense aware redundancy minimization process. On an average, for all 7 locations, we have observed 51.22% redundancy elimination done by the said process. So, in summary, we can conclude that our sense-aware redundancy elimination approach has successfully eliminated almost half of the redundancies present in the discovered activities. So, this analysis indirectly ensures that using `ActMiner`, we can reduce bandwidth wastage and push more unique information to the user while recommending in broadcast environment. The Roorkee dataset has also not been utilized in the said experimental analysis due to the same reason as discussed before.

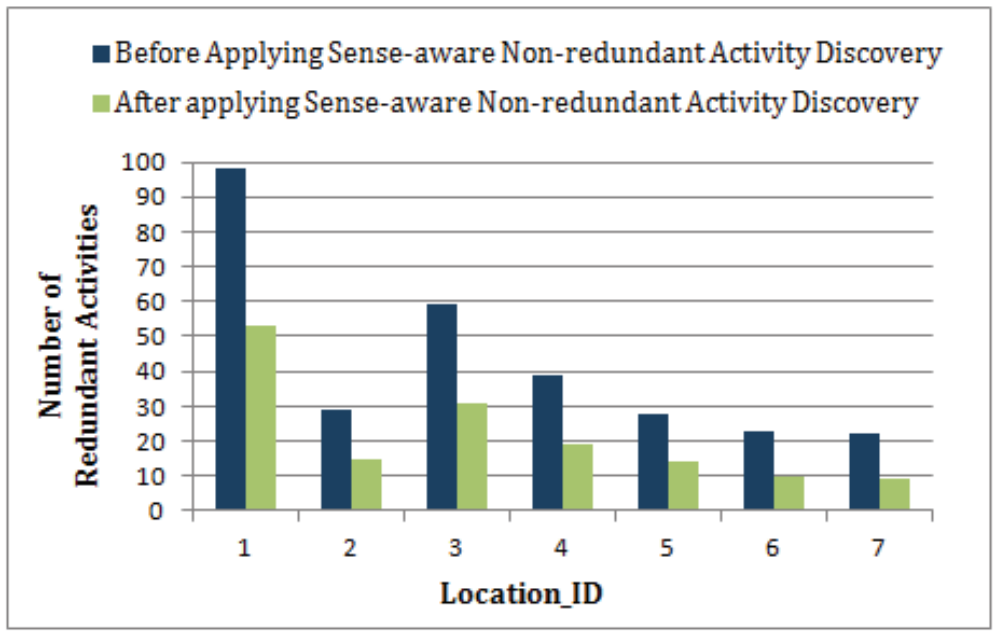

Figure 6.10: Performance of `ActMiner` in redundancy minimization

### 6.2.4   Evaluation 4: Qualitative Analysis II: Activity Ranking for Skyline Query Processing.

We have mentioned earlier that `LANet` can also be used for skyline query processing purpose. In skyline query processing, ranking of information is very important where the data that are not dominated by any other data, are retrieved from the database. Therefore, incorrect ranking of information in knowledgebase can cause retrieval of incorrect query results for recommendation purpose.

Besides addressing relevancy and redundancy issues, our proposed techniques category-aware relevant activity discovery and sense-aware non-redundant activity discovery of

`ActMiner` also cause improvement in activity ranking. Category-aware relevant activity discovery eliminates irrelevant activities and causes lower ranked activities to move upward in the rank list. Similarly, sense-aware non-redundant activity discovery merges two or more redundant activities into a single one and the individual frequencies of merged activities are add up and shifts the merged activity upward in the list. Moreover, due to merging the spaces occupied by activities that are being merged, become vacant and get occupied by next lower ranked activities which intern causes improvement in their ranks. In this way, we finally achieve an improved ranked list of activities after processing through successive stages of `ActMiner`.

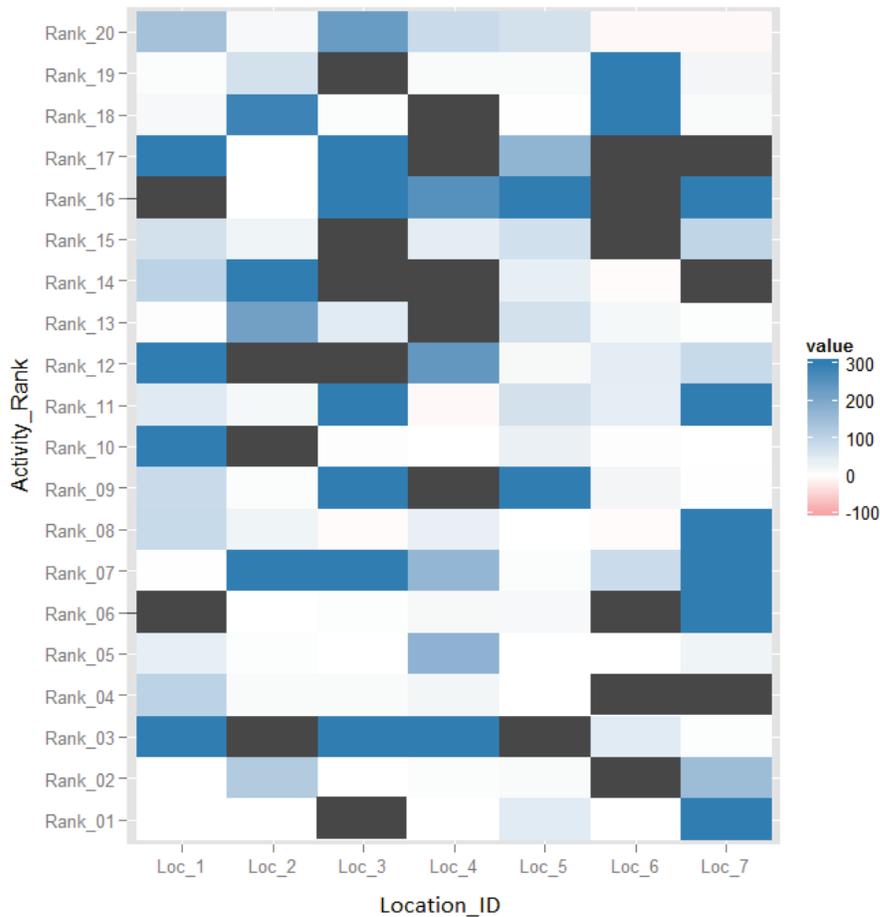

Figure 6.11: Heat map showing improvement in activity ranking on Yelp dataset.

In order to investigate the improvement in activity ranking, we have conducted our experiment in the following way. We first found out top 20 activities for each of the 7 yelp locations and then, discovered the rank of those activities in the rank list obtained using the baseline method. Next, we calculated the shifts in rank of each of those 20 activities by comparing the ranks of those activities in the inferred rank list obtained using baseline method and plotted the results in terms of a heat map as shown in the Figure 6.11. In the heat map, the blue colour shows the improvement in rank of activities. Black colour shows that the activity occupying the corresponding rank didn't appear in the rank list obtained using baseline approach and red colour has been used to show the degradation in rank of the activities in our top 20 activity rank list compared to that of the baseline method. In our experiment, we have obtained highest rank improvement of 4704 for activity rank 3 at location 1 and 1000+ rank improvements for 18 activity ranks considering all 7 locations. In the heat map shown above, we have scaled the map considering maximum rank improvement as 300 by reducing all 300+ activity rank improvements to 300 for better visualization. Apart from significant rank improvement in top-20 list of activities for each of the 7 locations, we have also obtained 25 new activities that appeared in our rank list but have not been inferred by the baseline approach in spite of the fact that they have high frequency which accounts for the reason behind acquiring a place by them in the top-20 activity list. The number of -ve improvement or degradation in the ranks are very less and insignificant compared to the +ve improvement as we can visualize in the heat map which places our knowledge base `LANet` one step forward compared to the baseline approach regarding the quality of information inferred by our system.

### 6.2.5 Evaluation 5: Qualitative Analysis III: Recommendation System.

To evaluate the performance in activity recommendation using `LANet`, we have developed a location-aware activity recommendation system with `LANet` as its back-end knowledgebase consisting of the activities inferred using `ActMiner`. Given an activity $A$ and a set of locations $Lset$, the recommender system recommends a location $L \in Lset$ such that $AF(A)$ is highest for location $L$. We have also developed similar recommendation system using the activities inferred by the baseline approach and evaluated both the recommender systems using "*Win-Loss Experiment*". In this evaluation, if the location IDs recommended by both the recommender systems are same, we have declared the result as "*Draw*". Otherwise, the recommender system which recommends the location with higher activity frequency value, wins in the experiment. We have discovered the set of distinct activities for for all locations in each data set and used them as a query input for the evaluation.

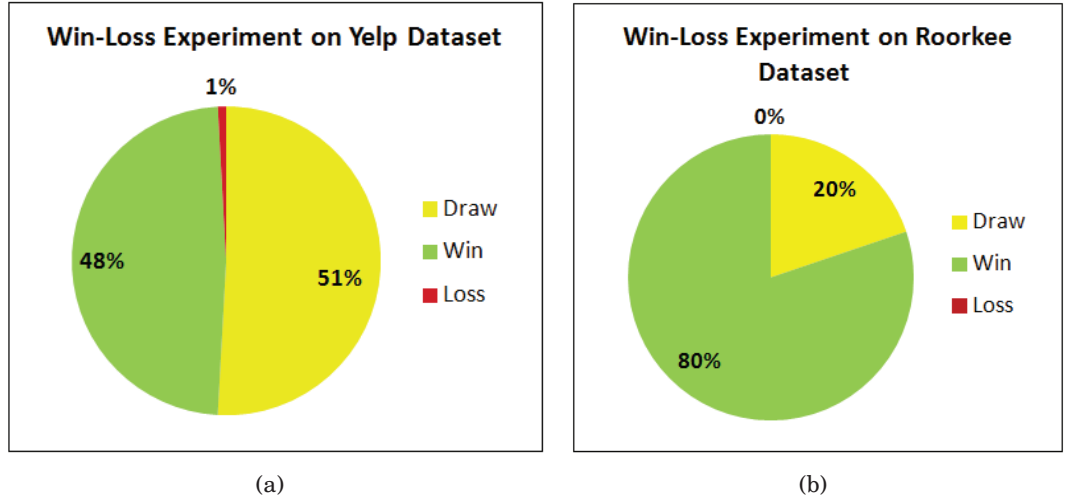

Figure 6.12: Result of Win-Loss Experiment on (a) Yelp data set and (b) Roorkee data set.

Figure 6.12(a) shows the comparative results of Win-Loss Experiments of the two recommender systems. Here, we observe that the `LANet`-based recommender system wins 48% cases and loses in only 1% case while competing with the recommender system formed using the baseline approach, whereas in 51% cases, the results are "*Draw*". Considering Roorkee dataset, `ActMiner`-based recommender system wins 80% cases and makes a draw in 20% cases without any loss as shown in the Figure 6.12(b). From these results, it is quite clear that the proposed knowledgebase `LANet` outperforms the baseline approach with respect to the performance recommendation which also proves the efficacy of our proposed solution.

## CONCLUSION AND FUTURE WORK

D ISCOVERING knowledge from data doesn't ensure its efficient usage. It's the efficient and systematic way of representing knowledge that makes it useful. In this dissertation, we have not only formulated the novel problem of discovering location-specific *relevant* and *non-redundant* activities from location-aware reviews, we have also developed a systematic way to represent the inferred knowledge in the form of an information netwrok- `LANet`. We have modeled the said information network in the form of a property graph that is stored in a graph database to serve as an efficient and enriched back-end knoledgebase of any location-aware activity recommendation system.

In order to meet our goal, we have introduced several novel techniques and ideas like `ActMiner` (for discovering location-specific *relevant* and *non-redundant* activities), *Activity-based Location Similarity Detection* (for alternate location recommnedation) and *Boundary of Uniqueness* (for quantifying the uniqueness property of an activity at a particular location) that played their respective roles in the enrichment process of the `LANet`. We have not only discovered and represented knowledge in an efficient and fruitful way, but also performed real-world experiments to justify our claims. The experimental analysis of the proposed knowledgebase has been performed to a reasonable depth which advocates for the accuracy and efficacy of our proposed solution.

## 7.1  Future Research Directions

There are several promising directions in which one can extend the work presented in this dissertation.

Firstly, we have built `LANet` based on only spatial aspect of the location-specific activities. We have not included the temporal aspect of activities performed at a particular location. Time management in today's busy world is an important aspect of life. People always try to devise ways for proper time utilization to shape their course of actions over time and space in their everyday lifestyle. Although with prior experience of performing an activity at a location, it becomes very easy to formulate the plan of visit and activity to be performed there; the situation becomes arduous if there is no such prior experience mostly encountered while visiting a new place. So, one possible extension of the proposed knowledgebase is to discover the suitable time of performing an activity at a given location and augment the `LANet` with these temporal information to enrich it further. Adding temporal dimension to `LANet` will make it enable to support activity-related temporal queries along with the spatial activity-related queries which `LANet` can support now.

Secondly, we have built the knowledgebase in the form of a property graph. But we haven't yet shown how we can build an *activity recommendation engine* that can interact with `LANet` for query processing purpose. In particular, we have not shown how the query specified by the end user in graphical interface will be mapped to the corresponding graph database query language like `Cypher(Neo4j)` to access and retrieve the query results from the database in real-time.

Last but not the least, we can make the knowledgebase adaptive such that it can be updated in real-time to keep up with the dynamic changes of locations and activities' popularity inferred from different Location-based social Networking platforms.